\documentclass[10pt, conference, compsocconf]{IEEEtran}

\usepackage{caption}
\usepackage{subcaption}

\usepackage{graphicx}
\usepackage{booktabs}
\usepackage{epstopdf}
\usepackage{float}
\usepackage{url}
\usepackage{placeins}
\usepackage{amsmath,amsfonts}
\usepackage{mathtools,mathrsfs}
\usepackage{multirow}
\usepackage{rotating}
\usepackage{pbox}
\usepackage[normalem]{ulem}

\usepackage{inconsolata}
\usepackage{listings}

\usepackage{cleveref}
\usepackage[utf8]{inputenc}
\crefname{section}{§}{§§}
\Crefname{section}{§}{§§}

\usepackage[10pt]{moresize}

\usepackage{xcolor}
\definecolor{darkgrey}{RGB}{70,70,70}
\definecolor{lightgrey}{RGB}{200,200,200}

\let\labelindent\relax
\usepackage{enumitem}

\usepackage{makecell}

\usepackage[linesnumbered,ruled,vlined]{algorithm2e}
\SetAlFnt{\scriptsize\tt}
\SetAlCapFnt{\scriptsize\sf}
\SetAlCapNameFnt{\scriptsize\sf}

\usepackage{algpseudocode}
\algblock{ParFor}{EndParFor}
\algnewcommand\algorithmicparfor{\textbf{parfor}}
\algnewcommand\algorithmicpardo{\textbf{do}}
\algnewcommand\algorithmicendparfor{\textbf{end\ parfor}}
\algrenewtext{ParFor}[1]{\algorithmicparfor\ #1\ \algorithmicpardo}
\algrenewtext{EndParFor}{\algorithmicendparfor}

\newcommand{\maciej}[1]{\textcolor{blue}{[Maciej: #1]}}

\newcommand{\goal}[1]{\noindent\textcolor{red}{[Goal: #1]}\par}

\newcommand{\nono}[1]{\textcolor{purple}{[Nono: #1]}}

\newcommand{\macb}[1]{\textbf{\textsf{#1}}}
\newcommand{\macbs}[1]{{\small\textbf{\textsf{#1}}}}

\usepackage[hang,flushmargin]{footmisc}

\usepackage{bm}

\usepackage{scalerel,stackengine}
\stackMath
\newcommand\rwh[1]{%
\savestack{\tmpbox}{\stretchto{%
  \scaleto{%
      \scalerel*[\widthof{\ensuremath{#1}}]{\kern-.6pt\bigwedge\kern-.6pt}%
          {\rule[-\textheight/2]{1ex}{\textheight}}
            }{\textheight}%
}{0.5ex}}%
\stackon[1pt]{#1}{\tmpbox}%
}

\def\HiLiGA{\leavevmode\rlap{\hbox to \hsize{\color{black!10}\leaders\hrule height 1\baselineskip depth 1ex\hfill}}}
\def\HiLiGB{\leavevmode\rlap{\hbox to \hsize{\color{black!25}\leaders\hrule height 1\baselineskip depth 1ex\hfill}}}
\def\HiLiGC{\leavevmode\rlap{\hbox to \hsize{\color{black!40}\leaders\hrule height 1\baselineskip depth 1ex\hfill}}}
\def\HiLiGD{\leavevmode\rlap{\hbox to \hsize{\color{black!55}\leaders\hrule height 1\baselineskip depth 1ex\hfill}}}
\def\HiLiGE{\leavevmode\rlap{\hbox to \hsize{\color{black!70}\leaders\hrule height 1\baselineskip depth 1ex\hfill}}}
\def\HiLiGF{\leavevmode\rlap{\hbox to \hsize{\color{black!85}\leaders\hrule height 1\baselineskip depth 1ex\hfill}}}

\usepackage{algpseudocode}
\usepackage{tikz}
\usetikzlibrary{calc}
\usepackage{xcolor}
\makeatletter

\renewcommand{\goal}[1]{}
\renewcommand{\maciej}[1]{}
\renewcommand{\nono}[1]{}
\newcommand{\edgar}[1]{{\color{magenta} edgar: #1}}
\renewcommand{\edgar}[1]{}

\usepackage{soul}

\usepackage{tikz}
\usetikzlibrary{tikzmark}

\tikzstyle{comment} = [draw, fill=blue!70, text=white, text width=3cm, minimum height=1cm, rounded corners, align=left, font=\scriptsize]
\tikzstyle{background_alg} = [draw, fill=blue!20, opacity=0.4, inner sep=4pt, rounded corners=2pt]

\usetikzlibrary{shapes}
\usetikzlibrary{plotmarks}
\usetikzlibrary{calc,fit}

\begin{document}

\title{SlimSell: A Vectorizable Graph Representation for Breadth-First Search}

\author{\IEEEauthorblockN{Maciej Besta$^1$$^\dagger$, Florian Marending$^1$$^\dagger$, Edgar Solomonik$^2$, Torsten Hoefler$^1$}
\IEEEauthorblockA{
  \textit{$^1$Department of Computer Science, ETH Zurich}\\
  \textit{$^2$Department of Computer Science, University of Illinois at Urbana-Champaign
  %
  %
  } \\
}
$^\dagger$Both authors contributed equally to this work.
}

\maketitle

\begin{abstract}
%
%
%
%
%
Vectorization and GPUs will profoundly change graph processing. Traditional graph
algorithms tuned for 32- or 64-bit based memory accesses will be inefficient on
architectures with 512-bit wide (or larger) instruction units that are already
present in the Intel Knights Landing (KNL) manycore CPU.
Anticipating this shift, we propose SlimSell: a vectorizable graph representation
to accelerate Breadth-First Search (BFS) based on sparse-matrix dense-vector
(SpMV) products.
%
%
%
SlimSell extends and combines the state-of-the-art SIMD-friendly Sell-$C$-$\sigma$ matrix
storage format 
%
%
with tropical, real, boolean, and
sel-max semiring operations. The resulting design reduces the
necessary storage (by up to 50\%) and thus pressure on the memory subsystem.
%
%
%
We augment SlimSell with the SlimWork and SlimChunk schemes that
reduce the amount of work and improve load balance, further
accelerating BFS.
%
%
We evaluate all the schemes on Intel Haswell multicore CPUs, the
state-of-the-art Intel Xeon Phi KNL manycore CPUs, and NVIDIA
Tesla GPUs. Our experiments indicate which semiring offers highest speedups for
BFS and illustrate that SlimSell accelerates a tuned Graph500 BFS code by up to
33\%.
%
%
%
This work shows that vectorization can secure high-performance in BFS based on
SpMV products; the proposed principles and designs can be extended to other
graph algorithms. 
%

\end{abstract}


{\vspace{0.5em}\noindent \textbf{This is an extended version of a paper published at\\ IEEE IPDPS'17 under the same title}}



\section{Introduction}


Upcoming CPUs armed with wide vector instruction units as well as GPUs with
even wider warps of cores will transform the way graph algorithms are reasoned
about, developed, and executed. Graph schemes are traditionally optimized for
architectures based on 32- or 64-bit instructions. This is not only due to the
prevalence of such systems but also because graph processing is most often
based on irregular fine-grained memory accesses~\cite{lumsdaine2007challenges, besta2017push}.
With the trend to develop machines deploying large vector instruction
or SIMD units, we must rethink the design of graph algorithms to use 
the available hardware effectively.

One such important algorithm is Breadth-First Search
(BFS)~\cite{Cormen:2001:IA:580470} that is used in many scientific domains such
as machine learning, data mining, and computational sciences. Moreover, it has
applications in various graph-related problems, including bipartiteness testing
and the Ford-Fulkerson method.
%
%
Thus, despite the wealth of existing optimizations for BFS, the growing amounts and
sizes of graph datasets require even faster BFS schemes.



It is well-known that BFS can be viewed in either the traditional combinatorial
or the algebraic abstraction. \emph{Traditional BFS} is based on primitives
such as queues. 
%
%
\emph{Algebraic BFS} is a
series of matrix-vector (MV) products over various semirings~\cite{kepner2011graph}; 
the products consist of a sparse matrix 
and a dense vector (SpMV) or a sparse matrix and a sparse vector (SpMSpV).
%
%
BFS based on SpMV (BFS-SpMV) uses no explicit locking~\cite{schmid2016high} or atomics~\cite{schweizer2015evaluating} and has a
succinct description as well as good locality~\cite{kepner2011graph, tate2014programming}.
Yet, it needs more work than traditional BFS and BFS based on
SpMSpV~\cite{yang2015fast}. 
%

\sethlcolor{while}
\begin{figure}[b]
  \centering
      \includegraphics[width=0.4\textwidth]{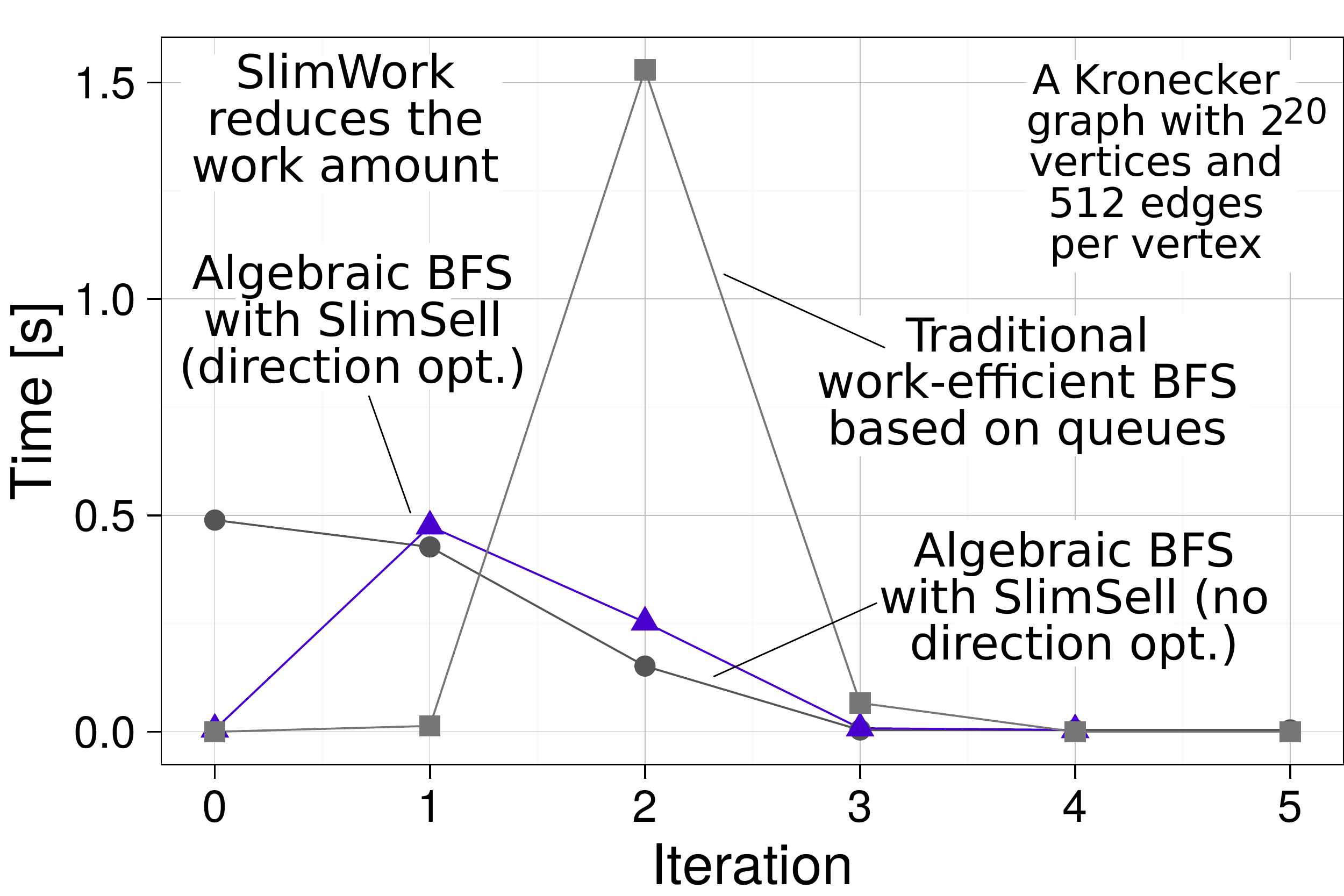}
      %
    \caption{Time to compute each BFS iteration for a Kronecker power-law graph on an Intel Xeon Phi Knights Landing manycore CPU.}
      \label{fig:motivate}
     \end{figure}
\sethlcolor{yellow}

There exist some BFS schemes for GPUs~\cite{paredes2016breadth,
merrill2015high, cheng2014understanding, yang2015fast, deng2009taming,
harish2007accelerating}.  However, they usually underutilize the available SIMD and
vectorization parallelism as they focus on work-optimal traditional BFS or BFS based on SpMSpV 
that use fine-grained irregular memory accesses. Moreover, they are
often tuned to a given architecture and thus not portable.
%
%
%
In this work, we illustrate that even if BFS-SpMV is not work-optimal,
one can use vectorization to outweigh potential
performance penalties and secure high-performance.
%
%
%
For this, we first show how to combine the
Sell-$C$-$\sigma$~\cite{DBLP:journals/corr/KreutzerHWFB13} sparse matrix
storage format with the algebraic BFS based on tropical, real, boolean, and
sel-max semirings. Intuitively, Sell-$C$-$\sigma$ decomposes a matrix into
\emph{chunks} stored in such a way that each chunk can be processed without
additional overheads by the whole SIMD execution unit. 
We conduct a work/storage complexity and performance analysis and discover which
semiring is the most advantageous for BFS-SpMV based on Sell-$C$-$\sigma$
for a large selection of parameters.
To the best of our knowledge, this is the first theoretical work complexity
analysis as well as performance study on the benefits and downsides
of each considered semiring. 

The advantage of \mbox{Sell-$C$-$\sigma$} is that it enables portable
performance for SpMV products on CPUs and GPUs.  Still, Sell-$C$-$\sigma$
was not originally built for graph processing where it does not perform well.
To alleviate this, we propose SlimSell: a vectorizable graph representation that
needs up to 50\% less storage than Sell-$C$-$\sigma$,
lowering the pressure on the memory subsystem when traversing undirected
graphs. Intuitively, we do not explicitly store non-zero entries of the
adjacency matrix but instead derive them implicitly from column indices.
We then augment SlimSell with the SlimWork and the SlimChunk schemes to reduce the amount of work
and load imbalance,
and thus accelerate BFS even further. Here, two key ideas are to: skip chunks
of computations if they correspond to final output values (e.g., vertex
distances) that would not change, and divide large chunks among multiple compute units;
an example illustration of SlimSell advantages is in Figure~\ref{fig:motivate}.
We then analyze the work and storage complexity of BFS based on SlimSell and derive bounds
for general graphs as well as for graph models that follow uniform (Erdős-Rényi)
  and power-law vertex degree distributions.
Our evaluation on multi- and manycore CPUs and GPUs shows that the resulting
BFS-SpMV outperforms traditional BFS by up to 33\%.
Finally, this is one of the first works to report BFS results on the state-of-the-art Intel
Knights Landing (KNL) Xeon Phi manycore.

We provide the following contributions:

\begin{itemize}[noitemsep, leftmargin=0.5em]
\item We combine the state-of-the-art Sell-$C$-$\sigma$ matrix storage format
with algebraic BFS based on tropical, boolean, real, and sel-max semirings.
We conduct work/storage complexity and performance analysis and show which
semiring secures the highest performance.
\item We introduce SlimSell, a vectorizable graph representation that improves the
storage complexity of Sell-$C$-$\sigma$ and accelerates BFS based on SpMV products.
\item We propose SlimWork and SlimChunk, two schemes that reduce the amount of work 
and improve load balancing in BFS
based on SlimSell.
\item We evaluate SlimSell and show performance
improvements of 33\% over the optimized Graph500 BFS code and storage reductions of 50\% 
over Sell-$C$-$\sigma$. Our work covers multi- and manycore CPUs as well as GPUs,
with \emph{the total of seven different systems};
we provide one of the first BFS results on the Intel KNL manycore.
\end{itemize}

\section{Background}


We start with presenting the utilized concepts.
%

\subsection{Notation}

For convenience, Table~\ref{tab:symbols} lists the most important symbols.

\subsubsection{Graph Structure}
We describe a graph $G$ as a tuple $(V,E)$; $V$ is a set of vertices and $E
\subseteq V \times V$ is a set of edges between vertices; $|V|=n$ and $|E|=m$.
$\mathbf{A}$ is the adjacency matrix of $G$. 
%
%
%
$G$'s diameter is denoted as $D$.
%
%
%


%


\subsubsection{Linear Algebra}
Bold symbols refer to vectors and matrices while standard font indicates a
vector/matrix element. The matrix-vector (MV) multiplications are denoted
with $\otimes$. The element-wise vector-vector (Hadamard) product is
referred to with $\odot$. The lower index is used to indicate the iteration
number (e.g., $\mathbf{x}_{k}$ is $\mathbf{x}$ in iteration~$k$) while the
upper index refers to a vector element (e.g., $x_k^s$ is the $s$th
element of $\mathbf{x}_k$); we also use square brackets when referring to
vector elements in the source code when they are implemented with arrays
($x_k^s \equiv x_k${\small\texttt{[$s$]}}).
The \emph{logical negation} of a vector $\mathbf{x}$ is an element-wise operation
$\overline{\mathbf{x}}$ where $\overline{x}^s = 0$ if $x^s \in \mathbb{R}
\setminus \{0\}$, and $\overline{x}^s = 1$ if $x^s = 0$ 
(e.g., $\overline{(0, 7, 0)^T} = (1, 0, 1)^T$).
%

\begin{table}
\centering
\scriptsize
\sf
\begin{tabular}{@{}l|ll@{}}
\toprule
\multirow{3}{*}{\begin{turn}{90}\shortstack{{Graph}\\{struct.}}\end{turn}} 
                    & $G$&A graph $G=(V,E)$; $V$ and $E$ are sets of vertices and edges.\\
                    & $n,m$&Numbers of vertices and edges in $G$; $|V| = n, |E| = m$.\\
                    & $\overline{\rho}, D$&Average degree and the diameter of $G$, respectively.\\
                   \midrule
\multirow{5}{*}{\begin{turn}{90}\shortstack{{Algebraic}\\{notation}}\end{turn}} 
                    & $\otimes, \odot$ & Matrix-vector (MV); element-wise vector-vector (VV) product.\\
                    & $\mathbf{A},\mathbf{A'}$& $G$'s adjacency matrix and the transformed $\mathbf{A}$ used in~\cref{sec:tropical}.\\
                    & $\mathbf{x}_k,\mathbf{f}_k$& The result of $\mathbf{A} \otimes \mathbf{f}_k$; the frontier of vertices in iteration~$k$.\\
                    & $\mathbf{d}, \mathbf{p}$ & Distances (to the root) and parents (in th BFS traversal tree). \\
                    & $DP(\mathbf{d})$ & A transformation that derives vertex parents $\mathbf{p}$ from $\mathbf{d}$. \\
                   \midrule
\multirow{3}{*}{\begin{turn}{90}\shortstack{Various}\end{turn}} 
                    & $T,W$ &The number of threads; work complexity of a given scheme.\\
                    & $C,n_c$ &Chunk height; the number of chunks in SlimSell or Sell-$C$-$\sigma$.\\
                    & $\sigma$ &Sorting scope in SlimSell and Sell-$C$-$\sigma$ ($\sigma \in [1,n]$).\\
\bottomrule
\end{tabular}
\caption{The most important symbols used in the paper.}
\label{tab:symbols}
\end{table}

\begin{lstlisting}[aboveskip=0em,float=h!,label=lst:vec-syntax,caption=Used vectorization functions: syntax and semantics., belowskip=-3em]
// |\underline{Notes:}| V is a vector of length $C$ of int: V $\equiv$ int[$C$].
// The names of the used |Intel functions| can be found in Listing |\ref{lst:haswell-syntax}|.

V LOAD(V* addr); // Load and return $C$ elements from $addr$.
void STORE(V* addr, V data); // Store $data$ at address $addr$.
V [$t_1$,$t_2$,...,$t_3$]; // Create a vector of $C$ elements.

// Compare a and b elementwise using operand $op$; return a vector 
// with binary outcome of each comparison (0/1 if 
// different/same); op can be EQ (equality), NEQ (non-equality).
V CMP(V a, V b, Op op);

// Interleave elements of a and b based on mask elements;
// |$\forall{i \in [0,C]}:$| $out[i]$ = $mask[i]$ ? $b[i]$ : $a[i]$
V BLEND(V a, V b, V mask);

// Return t elementwise result of the function FUN applied to 
// a and b. FUN can be: MIN (minimum), MAX (maximum), ADD (sum), 
// MUL (product), AND (logical and), OR (logical or).
V FUN(V a, V b);
\end{lstlisting}

\begin{figure*}
    \centering
    \includegraphics[width=1.0\textwidth]{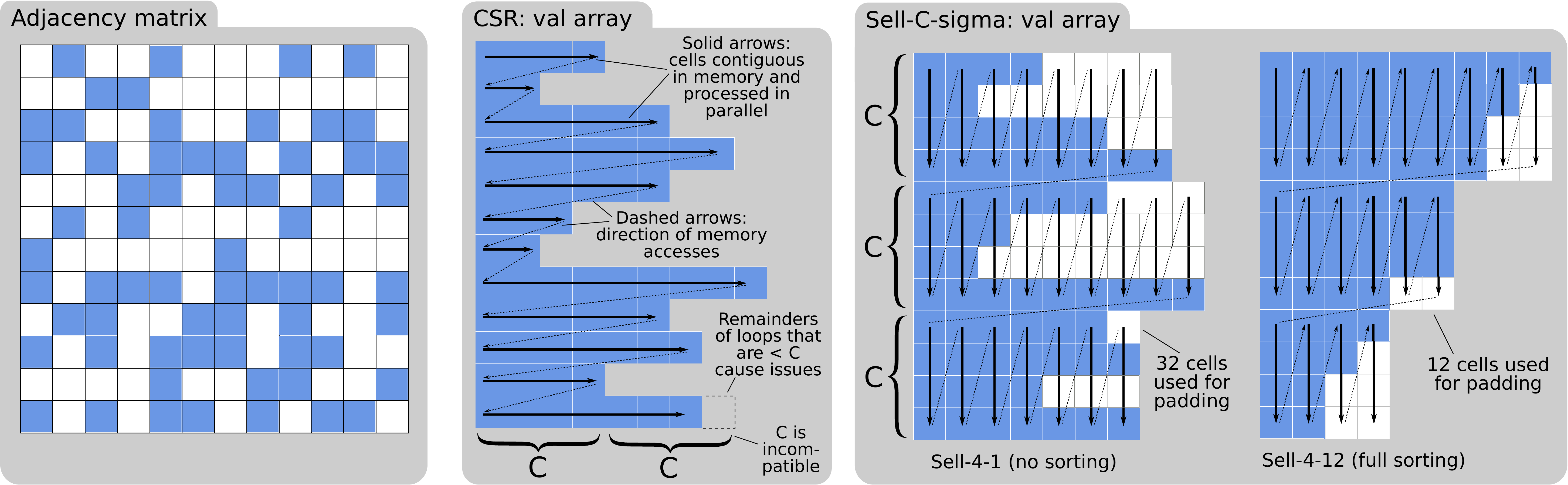}
    \caption{CSR and Sell-$C$-$\sigma$ constructed from an adjacency matrix. We only show $val$ arrays that illustrate the differences between representations.}
    \label{fig:csr-sell}
\end{figure*}

\subsection{Vectorization: Concepts and Considered Architectures}

Modern x86 CPUs feature single instruction, multiple data (SIMD) \emph{execution
units} (also called \emph{vector instruction units})~\cite{lomont2011introduction}.
%
%
One unit has several \emph{lanes} that proceed synchronously in
parallel and execute a specified scalar instruction on a data vector; each lane
processes a vector element of a specified size.
In modern CPUs with AVX (Advanced Vector Extensions), registers that hold the vectors are 256-bit wide, enabling 8
single-precision (or 4 double-precision) floating point operations per
instruction. Consequently, the \emph{SIMD width} is respectively 8 and~4.
%
%
%
%
Next, manycores such as Xeon Phi KNL use wider SIMD units; KNL comes with 
512-bit wide units.
Finally, GPUs employ SIMT where \emph{CUDA cores} roughly correspond
to SIMD lanes. They are scheduled in \emph{warps}; one warp usually counts 32
cores, which constitutes the GPU ``SIMD width''.
%
%
We present the used vectorization functions in Listing~\ref{lst:vec-syntax}.
For a widespread analysis of SlimSell, we evaluate all three types of architectures: 
Intel CPUs (represent classic latency-oriented x86 multicores),
Intel Xeon Phi KNL (stands for state-of-the-art manycores), and NVIDIA GPUs
(represent modern throughput-oriented SIMT GPUs).

\begin{lstlisting}[label=lst:haswell-syntax, belowskip=0em, caption=The
syntax/semantics of the utilized vector functions on Intel x86]
/* |\underline{Notes:}| T is an arbitrary type supported by the
 * AVX instruction set extension. C is determined 
 * by the size of one instance of T: an AVX register
 * can hold $C=256 / (8*sizeof(T))$ elements of type T.
 * 
 * 'vec' describes a vector of $C$ elements of type T and is
 * stored in a 256-bit AVX register.
 *
 * Following the simplified syntax we provide the real
 * instruction names as specified by the AVX instruction set
 * extension. For this we assume T to be 32-bit floating
 * point. For different types the names vary. */

/* Load C elements at address $addr$ into a vector. */
vec load(T * addr);
__m256 _mm256_load_ps(float const * addr)

/* Store the C elements contained in $data$
 * at address $addr$. */
void store(T * addr, vec data);
void _mm256_store_ps(float * addr, __m256 data);

/* Create a vector of length C from C individual elements */
vec set(T $t_C$, ... , T $t_1$);
__m256 _mm256_set_ps(float t_8, float t_7, ... , float t_1);

/* Create a vector of C repeated instances of $t$ */
vec set1(T $t$);
__m256 _mm256_set1_ps(float a);

/* Compare vectors a and b elementwise based on comparison
 * operand $op$ and return a vector containing binary
 * information on the outcome of each comparison. */
vec cmp(vec a, vec b, Op op);
__m256 _mm256_cmp_ps(__m256 a, __m256 b, const int op);

/* Interleave elements of vectors a and b
 * based on corresponding element
 * of mask: |$\forall{i \in [0,C]}:$| $out[i]$ = $mask[i]$ ? $b[i]$ : $a[i]$ */
vec blend(vec a, vec b, vec mask);
__m256 _mm256_blend_ps(__m256 a, __m256 b, const int mask);

/* Convert vector a of integer type to floating point type */
vec cvtI2f(vec a);
__m256 _mm256_cvtepi32_ps(__m256i a);

/* Return a vector containg the elementwise minimum of the
 * vectors a and b. */
vec min(vec a, vec b);
__m256 _mm256_min_ps(__m256 a, __m256 b);

/* Return a vector containg the elementwise maximum of the
 * vectors a and b. */
vec max(vec a, vec b);
__m256 _mm256_max_ps(__m256 a, __m256 b);

/* Return a vector containg the elementwise addition of the
 * vectors a and b. */
vec add(vec a, vec b);
__m256 _mm256_add_ps(__m256 a, __m256 b);

/* Return a vector containg the elementwise multiplication of
 * the vectors a and b. */
vec mul(vec a, vec b);
__m256 _mm256_mul_ps(__m256 a, __m256 b);

/* Return a vector containg the elementwise logical and of
 * the vectors a and b. */
vec and(vec a, vec b);
__m256 _mm256_and_ps(__m256 a, __m256 b);

/* Return a vector containg the elementwise logical or of the
 * vectors a and b. */
vec or(vec a, vec b);
__m256 _mm256_or_ps(__m256 a, __m256 b);

\end{lstlisting}

%

%

%


\subsection{Algorithms: BFS in Various Abstractions}
\label{sec:bfs-background}



The goal of BFS is to visit each vertex in $G$. First, it visits the neighbors
of a specified \emph{root} vertex $r$, then all the unvisited
neighbors of $r$'s neighbors, and so on. Vertices that are
processed in a current step constitute the BFS \emph{frontier}.
BFS outputs either the distance (in hops) from each vertex to
$r$, or the parent of each vertex in the BFS traversal tree.
Distances are stored in a vector $\mathbf{d} \in \mathbb{R}^n$ ($d^v$ is the
distance from vertex~$v$ to $r$) and parents in $\mathbf{p} \in
\mathbb{R}^n$ ($p^v$ is the parent of vertex~$v$).
$\mathbf{d}$ can be transformed into $\mathbf{p}$ using $O(m+n)$ work and
$O(1)$ depth. To achieve this, for each vertex $v$, the neighbor $w$ of $v$ with
the distance $d^w = d^v - 1$ must be found;
%
%
%
we refer to it as the \emph{DP transformation} ($\mathbf{p} =
DP(\mathbf{d})$).
%
%
%

%
%
%

\subsubsection{Traditional BFS}

Here, a frontier is implemented with a queue $Q$.  At every iteration, vertices
are removed in parallel from $Q$ and all their unvisited neighbours are
appended to $Q$; this process is repeated until $Q$ is empty.

\subsubsection{Algebraic BFS}

BFS can also be implemented with the MV product over a selected semiring.
%
%
For example, for the real semiring, $\mathbf{x}_0 \in \mathbb{R}^n$ is a starting vector with $x_0^r = 1$ and
$0$ elsewhere. 
Then, for iteration~$k \ge 1$, $\mathbf{x}_k = \mathbf{A}
\otimes \mathbf{f}_{k-1}$ (we assume that $\mathbf{A}$ is pre-transposed in
accordance with common practice~\cite{bulucc2011parallel});
$\mathbf{f}_{k-1}$ is the frontier in iteration $k-1$; note that $\mathbf{f}_0
= \mathbf{x}_0$.
%
%
As we show in~\cref{sec:accelerate_bfs_variants}, for some semirings
$\mathbf{f}_k = \mathbf{x}_k$,
while for others one must derive $\mathbf{f}_k$
from $\mathbf{x}_k$.
$\mathbf{x}$ and $\mathbf{f}$ can be represented either as sparse or dense
vectors that respectively result in sparse-sparse and sparse-dense $\mathbf{A} \otimes \mathbf{f}$
products; the latter entail more work but offer more potential for
vectorization that we use in SlimSell.


%

\subsection{Data Structures: CSR, Sell-$C$-$\sigma$, and Adjacency List}


Finally, we discuss data structures relevant for our work.

\subsubsection{Compressed Sparse Row (CSR)}
\label{sec:csr_back}


%
In the well-known CSR format (see Figure~\ref{fig:csr-sell}), a matrix is represented with three arrays: $val$,
$col$, and $row$. $val$ contains all $\mathbf{A}$'s non-zeros (that correspond to $G$'s edges) in the row major
order. $col$ contains the column index for each corresponding value in $val$
and thus has the same size ($O(m)$). Finally, $row$ contains starting indices in $val$ (and
$col$) of the beginning of each row in $\textbf{A}$ (size $O(n)$).
%
%
CSR is widely adopted for its simplicity and low memory footprint for sparse
matrices. Yet, it hinders high-performance vectorization.
Consider the CSR-based textbook MV product in Listing~\ref{lst:mvm-csr-unroll}
where we unroll the inner loop.  Such unrolling ensures consecutive data
accesses necessary to effectively use vectorization (consecutive $val$ entries
are exposed). Yet, due to the reduction
(line~\ref{line:vector-reduction}) and the cleanup of the loop remainder
(line~\ref{line:loop-remainder}), the unrolling becomes inefficient for
sparse matrices~\cite{DBLP:journals/corr/KreutzerHWFB13}. 




\begin{lstlisting}[aboveskip=-0.5em,float=h!,label=lst:mvm-csr-unroll,caption=(\cref{sec:csr_back}) MV 
product with CSR and 4-way loop unrolling. Data alignment is assumed for clarity.]
/* The code computes $\mathbf{x}_{k+1} = \mathbf{A} \otimes \mathbf{x}_k$ using CSR. */
for(int i = 0; i < $n$; i++) {
  int t_0 = t_1 = t_2 = t_3 = 0;
  for(int j = $row$[i]; j < $row$[i+1]; j += 4) {
    t_0 += $val$[j+0]*$x_{k}$[$col$[j+0]]; t_1 += $val$[j+1]*$x_{k}$[$col$[j+1]];
    t_2 += $val$[j+2]*$x_{k}$[$col$[j+2]]; t_3 += $val$[j+3]*$x_{k}$[$col$[j+3]];
  }
  $x_{k+1}$[i] = t_0 + t_1 + t_2 + t_3; // Reduction.|\label{line:vector-reduction}|
  // Clean up the loop remainder:
  for(j=j-4; j < $row$[i+1]; j++) {$x_{k+1}$[i] += $val$[j]*$x_{k}$[$col$[j]];}}|\label{line:loop-remainder}|
\end{lstlisting}

%

\begin{lstlisting}[belowskip=-3em,aboveskip=0.5em,mathescape,float=h!,label=lst:mvm-sell,caption=(\cref{sec:sell_back}) The MV product 
kernel with Sell-$C$-$\sigma$ ($C$ is $4$).]
/* The code computes $\mathbf{x}_{k+1} = \mathbf{A} \otimes \mathbf{x}_k$ using Sell-$C$-$\sigma$. */
for(int i = 0; i < $n$/$C$; ++i) {
  for(int j = 0; j < $cl$[i]; ++j) {
    $x_{k+1}$[i*$C$+0] += $val$[$cs$[i]+j*$C$+0] * $x_{k}$[$col$[$cs$[i]+j*$C$+0]];
    $x_{k+1}$[i*$C$+1] += $val$[$cs$[i]+j*$C$+1] * $x_{k}$[$col$[$cs$[i]+j*$C$+1]];
    $x_{k+1}$[i*$C$+2] += $val$[$cs$[i]+j*$C$+2] * $x_{k}$[$col$[$cs$[i]+j*$C$+2]];
    $x_{k+1}$[i*$C$+3] += $val$[$cs$[i]+j*$C$+3] * $x_{k}$[$col$[$cs$[i]+j*$C$+3]]; } }
\end{lstlisting}

\subsubsection{Sell-$C$-$\sigma$ Format}
\label{sec:sell_back}


To overcome the issues in CSR, Kreutzer et al.\ proposed a sparse matrix
format called Sell-$C$-$\sigma$ that exposes data parallelism portably
across architectures with different SIMD
widths~\cite{DBLP:journals/corr/KreutzerHWFB13}. 
Intuitively, the key move from CSR is to ``turn by 90$^{\circ}$'' the layout
of $val$ and $col$ in memory so that \emph{consecutive SIMD lanes/cores can process
in parallel consecutive {rows} of $\mathbf{A}$}.
We illustrate Sell-$C$-$\sigma$ in Figure~\ref{fig:csr-sell} and
Listing~\ref{lst:mvm-sell}. It has four arrays: $val$, $col$, $cs$, and $cl$.
Like in CSR, $val$ and $col$ host non-zeros in $\mathbf{A}$ and the
related column indices.
The difference is that they consist of contiguous \emph{chunks}; each chunk
contains $C$ matrix rows. Chunk elements are stored in the \emph{column major}
order; chunks are consecutive in memory. The $cs$ and $cl$ arrays provide
starting offsets and lengths of each chunk. Rows are zero-padded up to the
longest row in a chunk. 
%
%
Similar-sized rows are stored together by sorting them by the length,
reducing the amount of required padding. Finally, $\sigma \in [1,n]$ 
controls the sorting scope; a larger $\sigma$ entails more sorting. 
In this work, we show how to adapt Sell-$C$-$\sigma$ to BFS
and to improve its design for more performance.

\subsubsection{Adjacency List (AL)}

Finally, AL is the well-known graph representation that uses $2m + n$ memory cells to store
an undirected and unweighted graph; it consists of an array with 
IDs of neighbors of each vertex (size $2m$) and an offset array with the
beginning of the neighbor data of each vertex (size $n$). AL is used together with traditional BFS.
We show that SlimSell has comparable or smaller size than AL and secures more
performance for various types of graphs.




\section{Accelerating BFS based on SpMV}

We now proceed to describe how to amortize and outweigh the inherent additional
work in BFS-SpMV using vectorization and SIMD parallelism.
The first key idea is to combine Sell-$C$-$\sigma$ with semirings to portably
accelerate BFS-SpMV (\cref{sec:accelerate_bfs_variants}).
The second core idea is to develop SlimSell, a vectorizable graph representation
that builds upon Sell-$C$-$\sigma$ by making it more storage-efficient
(\cref{sec:slimsell}). Finally, we propose SlimWork (\cref{sec:slimwork}) and
SlimChunk (\cref{sec:slimchunk}), two schemes that reduce the work amount and
improve load balancing in BFS-SpMV based on SlimSell. 





\subsection{Combining Sell-$C$-$\sigma$ with Semirings}
\label{sec:accelerate_bfs_variants}

We now 
provide \emph{the first systematic comparison of 
semirings for BFS-SpMV},
and combine them with
Sell-$C$-$\sigma$; see 
Listing~\ref{lst:bfs-spmv-semirings}
for
a BFS iteration (frontier expansion) with the considered semirings. 
A reader who is not interested in the following theoretical details can proceed directly to~\cref{sec:slimsell}.
A semiring is defined as a tuple $S=(X, op_1, op_2, el_1, el_2)$ 
where $X$ is a set equipped with two binary operations $op_1, op_2$ such that $(X, op_1)$
is a commutative monoid with identity element $el_1$ and $(X, op_2)$ is a monoid
with identity element $el_2$.
We denote matrix-vector multiplication and Hadamard products on semiring $S$ as $\otimes_S$ and $\odot_S$.
%

\subsubsection{The Tropical Semiring $T=(\mathbb{R} \cup \{\infty\}, \min, +, \infty, 0)$}
\label{sec:tropical}

To use BFS with this semiring, one must first transform $\mathbf{A}$ to $\mathbf{A'}$ where
each off-diagonal zero entry is $\infty$. Next, one sets ${x}_0^r = 0$ and ${x}_0^s =
\infty$, for $s \neq r$ (the initial distance to $r$ and any other vertex is
respectively $0$ and $\infty$).  Then, $\mathbf{x}_k = \mathbf{f}_k =
\mathbf{A'} \otimes_T \mathbf{f}_{k-1}$.
This BFS computes distances \emph{after the final
iteration}: $\mathbf{d} = \mathbf{x}_{D}$ ($D$ is the diameter) with predecessors
$\mathbf{p} = DP(\mathbf{d})$ (see \cref{sec:bfs-background} for a description of $DP$).
%

%


\begin{tikzpicture}[remember picture,overlay,pin distance=0cm]
\draw[fill=black!10, opacity=1, inner sep=4pt, rounded corners=2pt]
  ([shift={(-0.25em,-10.25em)}]pic cs:bfs-spmv-tr-1-s)
    rectangle
  ([shift={(23.5em,-11.7em)}]pic cs:bfs-spmv-tr-1-e);
\fill ([shift={(20.4em,-11.05em)}]pic cs:bfs-spmv-tr-1-e) node[rounded corners, text=white, fill=black!75, font=\tiny] {\macb{TROPICAL SEMIRING}};
\draw[fill=black!10, opacity=1, inner sep=4pt, rounded corners=2pt]
  ([shift={(-0.25em,-11.7em)}]pic cs:bfs-spmv-bl-1-s)
    rectangle
  ([shift={(23.5em,-13.1em)}]pic cs:bfs-spmv-bl-1-e);
\fill ([shift={(20.4em,-12.45em)}]pic cs:bfs-spmv-bl-1-e) node[thin,rounded corners, text=white, fill=black!75, font=\tiny] {\macb{BOOLEAN SEMIRING}};
\draw[fill=black!10, opacity=1, inner sep=4pt, rounded corners=2pt]
  ([shift={(-0.25em,-13.1em)}]pic cs:bfs-spmv-bl-1-s)
    rectangle
  ([shift={(23.5em,-14.5em)}]pic cs:bfs-spmv-bl-1-e);
\fill ([shift={(20.52em,-13.82em)}]pic cs:bfs-spmv-bl-1-e) node[thin,rounded corners, text=white, fill=black!75, font=\tiny] {\macb{SEL-MAX SEMIRING}};
\draw[fill=black!10, opacity=1, inner sep=4pt, rounded corners=2pt]
  ([shift={(-0.25em,-17.25em)}]pic cs:bfs-spmv-bl-1-s)
    rectangle
  ([shift={(23.5em,-18.7em)}]pic cs:bfs-spmv-bl-1-e);
\fill ([shift={(20.4em,-18em)}]pic cs:bfs-spmv-tr-1-e) node[rounded corners, text=white, fill=black!75, font=\tiny] {\macb{TROPICAL SEMIRING}};
\draw[fill=black!10, opacity=1, inner sep=4pt, rounded corners=2pt]
  ([shift={(-0.25em,-18.7em)}]pic cs:bfs-spmv-bl-1-s)
    rectangle
  ([shift={(23.5em,-26.4em)}]pic cs:bfs-spmv-bl-1-e);
\fill ([shift={(20.4em,-19.4em)}]pic cs:bfs-spmv-bl-1-e) node[thin,rounded corners, text=white, fill=black!75, font=\tiny] {\macb{BOOLEAN SEMIRING}};
\draw[fill=black!10, opacity=1, inner sep=4pt, rounded corners=2pt]
  ([shift={(-0.25em,-26.4em)}]pic cs:bfs-spmv-bl-1-s)
    rectangle
  ([shift={(23.5em,-33.5em)}]pic cs:bfs-spmv-bl-1-e);
\fill ([shift={(20.52em,-27.2em)}]pic cs:bfs-spmv-bl-1-e) node[thin,rounded corners, text=white, fill=black!75, font=\tiny] {\macb{SEL-MAX SEMIRING}};
\end{tikzpicture}

\begin{lstlisting}[float=h!,label=lst:bfs-spmv-semirings,caption=One iteration (frontier expansion) of BFS-SpMV; $C$ is $4$.
We omit the real semiring as it is very similar to the boolean one.]
// $val$, $col$, $cs$, $cl$ are data structures in Sell-$C$-$\sigma$ (see |\cref{sec:sell_back}|);

for(int i = 0; i < $n_c$; i++) { // Iterate over every chunk.
  int index = $cs$[i]; // Get offset to chunk i.
  V x = LOAD(&${f}_{k-1}$[i*$C$]); // Load chunk i from frontier ${f}_{k-1}$.
  for(int j = 0; j < $cl$[i]; j++) { // Iterate over each column. |\label{line:mv-inner-start}|
    V vals = LOAD(&$val$[index]); // Load $\mathbf{A'}$ values. |\label{line:load-val}|
    // Create the rhs vector consisting of $C$ $\mathbf{f}_{k-1}$ entries.
    V rhs = [${f}_{k-1}$[$col$[index+3]], ${f}_{k-1}$[$col$[index+2]],
             ${f}_{k-1}$[$col$[index+1]], ${f}_{k-1}$[$col$[index+0]]];

    // Compute $\mathbf{x}_{k}$ (versions differ based on the used semiring):
|\tikzmark{bfs-spmv-tr-1-s}|#ifdef USE_TROPICAL_SEMIRING    
    x = MIN(ADD(rhs, vals), x); |\tikzmark{bfs-spmv-tr-1-e}|
|\tikzmark{bfs-spmv-bl-1-s}|#elif defined USE_BOOLEAN_SEMIRING 
    x = OR(AND(rhs, vals), x);|\tikzmark{bfs-spmv-bl-1-e}|
|\tikzmark{bfs-spmv-sl-1-s}|#elif defined USE_SELMAX_SEMIRING 
    x = MAX(MUL(rhs, vals), x);|\tikzmark{bfs-spmv-sl-1-e}|
#endif
    index += $C$; 
  }|\label{line:mv-inner-end}|
  // Now, derive $\mathbf{f}_{k}$ (versions differ based on the used semiring):|\label{line:mv-post-start}|
|\tikzmark{bfs-spmv-tr-2-s}|#ifdef USE_TROPICAL_SEMIRING  
  STORE(&${f}_k$[i*$C$], x); // Just a store. |\tikzmark{bfs-spmv-tr-2-e}|
|\tikzmark{bfs-spmv-bl-2-s}|#elif defined USE_BOOLEAN_SEMIRING
  // First, derive $\mathbf{f}_k$ using filtering.
  V g = LOAD(&$g_{k-1}$[i*$C$]); // Load the filter $\mathbf{g}_{k-1}$.
  x = CMP(AND(x, g), [0,0,...0], NEQ); STORE(&$x_k$[i*$C$], x);

  // Second, update distances $\mathbf{d}$; depth is the iteration number.
  V x_mask = x; x = MUL(x, [depth,...,depth]); 
  x = BLEND(LOAD(&$d$[i*$C$]), x, x_mask); STORE(&$d$[i*$C$], x);

  // Third, update the filtering term.
  g = AND(NOT(x_mask), g); STORE(&$g_k$[i*$C$], g); |\tikzmark{bfs-spmv-bl-2-e}|
|\tikzmark{bfs-spmv-sl-2-s}|#elif defined USE_SELMAX_SEMIRING:
  // Update parents.
  V pars = LOAD(&$p_{k-1}$[i*$C$]); // Load the required part of $\mathbf{p}_{k-1}$
  V pnz = CMP(pars, [0,0,...,0], NEQ);
  pars = BLEND([0,0,...,0], pars, pnz); STORE(&$p_{k}$[i*$C$], pars);

  // Set new $\mathbf{x}_k$ vector.
  V tmpnz = CMP(x, [0,0,...,0], NEQ);
  x = BLEND(x, &$v$[i*$C$], tmpnz); STORE(&$x_k$[i*C], x); |\tikzmark{bfs-spmv-sl-2-e}|
#endif |\label{line:mv-post-end}|
\end{lstlisting}


\subsubsection{The Real Semiring $R=(\mathbb{R}, +, \cdot, 0 , 1)$}
\label{sec:r-semiring}

Here, ${x}_0^r = {f}_0^r = 1$ and any other element of $\mathbf{x}_0$ and
$\mathbf{f}_0$ equals 0 (i.e., only $r$ is in the initial frontier).  Then,
$\mathbf{x}_k = \mathbf{A} \otimes_R \mathbf{f}_{k-1}$ and $\mathbf{f}_k =
\mathbf{x}_k \odot_R \mathbf{g}_k$ where $\mathbf{g}_k$ is the \emph{filtering
term}: $\mathbf{g}_k = \overline{\sum_{l=0}^{k-1} \mathbf{f}_l}$.  Finally,
$\mathbf{d} = \sum_{l=1}^{D} l \mathbf{f}_l$ and $\mathbf{p} = DP(\mathbf{d})$.
This semiring results in code snippets similar to the following
boolean semiring and we omit it due to space constraints.

\subsubsection{The Boolean Semiring $B=(\{0,1\}, |, \&, 0, 1)$}

Similarly to the real semiring, ${x}_0^r = {f}_0^r = 1$ and 
${x}_0^s = {f}_0^s = 0, s \neq r$. 
%
%
Next, $\mathbf{x}_k = \mathbf{A} \otimes_B \mathbf{f}_{k-1}$, and
$\mathbf{f}_k = \mathbf{x}_k - \mathbf{g}_k$; $\mathbf{g}_k = \overline{\bigcup_{l=0}^{k-1} \mathbf{f}_l}$, $k
\ge 2$.
Then, $\mathbf{d} = \bigcup_{l=1}^{D} l \mathbf{f}_l$ and $\mathbf{p} = DP(\mathbf{d})$.

%
%
%
%


\subsubsection{The Sel-Max (S) Semiring $(\mathbb{R}, \max, \cdot, -\infty, 1)$}
All the above BFS variants require the $DP$ transformation to derive
$\mathbf{p}$. We now present a semiring 
that facilitates computing parents. In this semiring
$\mathbf{x}_0^r = r$ and $\mathbf{x}_0^s = 0, s \neq r$.
The frontier is given by $\mathbf{f}_0^r = 1$ and $\mathbf{f}_0^s = 0, s \neq r$.
Here, we also use an additional vector $\mathbf{p}_0 = \mathbf{x}_0$.
Then, in each iteration, we first calculate $\mathbf{x}_k = \mathbf{A} \otimes
\mathbf{x}_{k-1}$ and derive the current frontier $\mathbf{f}_k =
\overline{\overline{\mathbf{x}_{k}}} - \mathbf{g}_k$ where $\mathbf{g}_k = \sum_{l=0}^{k-1} \mathbf{f}_l$ 
($\overline{\overline{\mathbf{x}}}$ is a double logical negation to transform each non-zero in $\mathbf{x}$ into $1$).
\maciej{0-1} We next obtain $\mathbf{p}_k = \mathbf{p}_{k-1}
+ \overline{\mathbf{p}_{k-1}} \odot \mathbf{x}_k$. Moreover, $\mathbf{x}_k$
must be updated in such a way that each non-zero element becomes equal to its
index: $\mathbf{x}_k = \overline{\overline{\mathbf{x}_k}} \odot (1, 2, ...
n)^T$.  After the last iteration, we have $\mathbf{p} = \mathbf{p}_D$. Finally,
$\mathbf{d} = \bigcup_{l=1}^{D} l \mathbf{f}_l$.

\macbs{Work Complexity}
We compare work complexity $W$ of BFS-SpMV based on Sell-$C$-$\sigma$ with
other schemes in Table~\ref{tab:compar}.
Full sorting that takes $O(n \log n)$ is assumed ($\sigma = n$).
Still, we exclude this term as it
is a one-time investment and it does not change the asymptotic complexity.
Given vertices sorted in increasing order by degree, we let $\rho_i$ denote the degree of the $i$th 
vertex and $\rwh{\rho}$ denote the maximum vertex degree.
%
Our results focus on the tropical semiring, results for other semirings follow similarly. 
We derive a general bound for $W$ based only on $C$ and the maximum degree $\rwh{\rho}$ for any graph.
We then apply this bound to obtain expected storage bounds for Erdős-Rényi and power-law graph models.

\begin{table}[h!]
\footnotesize
\sf
\centering
\begin{tabular}{l|l}
\toprule
BFS algorithm & $W$ \\
\midrule
Traditional BFS (textbook~\cite{Cormen:2001:IA:580470})  & $O(n + m)$ \\
Traditional BFS (bag-based~\cite{leiserson2010work})  & $O(n + m)$ \\
Traditional BFS (direction-inversion~\cite{harish2007accelerating})  & $O(Dn + Dm)$ \\
BFS-SpMV (textbook~\cite{Cormen:2001:IA:580470})  & $O(D n^2)$ \\
BFS-SpMV~\cite{deng2009taming}  & $O(Dn + Dm)$ \\
BFS (SpMSpV with merge sort~\cite{yang2015fast}) & $O(n + m \log m)$ \\
BFS (SpMSpV with radix sort~\cite{yang2015fast})  & $O(n + xm)$ \\
BFS (SpMSpV with no sort~\cite{yang2015fast})  & $O(n + m)$ \\
\midrule
\textbf{This work} (graphs with max degree $\rwh{\rho}$)  & $O(Dn + Dm +DC\rwh{\rho})$ \\
\textbf{This work} (Erdős-Rényi graphs)  & See Equation~(\ref{eqn:W_ER}) \\
\textbf{This work} (power-law graphs)  & See Equation~(\ref{eqn:W_K})\\
\bottomrule
\end{tabular}
\caption{Comparison of $W$; $x$ is the length of the largest key in binary~\cite{yang2015fast}.}
\label{tab:compar}
\end{table}

\begin{figure}
    \centering
    \includegraphics[width=0.45\textwidth]{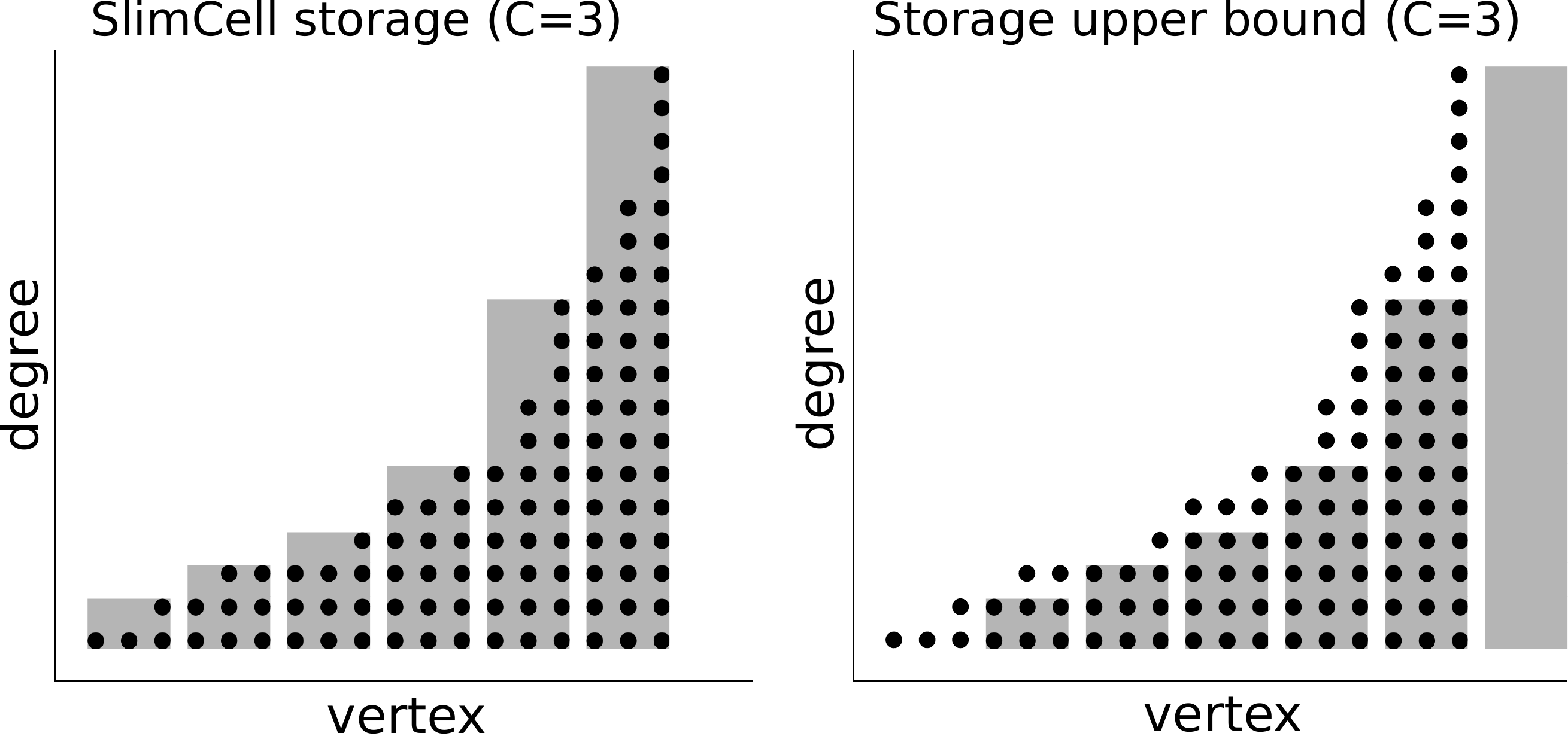}
    \caption{Justification of the upper bound on work and storage in Sell-$C$-$\sigma$ (for $C = 3$). 
             Overall, the padding can generate at most $C\hat{\rho}$ extra storage where $\hat{\rho}$ 
            is the maximum degree.}
    \label{fig:bound}
\end{figure}

The size of all the blocks (except the largest) is in total
$$\sum_{i=2}^{n_c} C\rho_{iC-1} \leq \sum_{i=1}^{n_c-1} \sum_{j=1}^C \rho_{(i-1)C+j}\leq m,$$
as the size of each block is smaller than the number of vertices in the previous (larger) block.
Next, the size of the largest block is $\rwh{\rho}C$, so the total storage is bounded by
$$\sum_{i=1}^{n_c} C\rho_{iC-1} \leq  m+\rwh{\rho}C.$$
This bound is illustrated in Figure~\ref{fig:bound}.
Asymptotically, this bound is always tight for Sell-$C$-$\sigma$, since the minimum storage (a lower bound)
is $\max(m,\rwh{\rho}C)$.

To obtain a bound on $W$, we assume all vertices and edges need to be accessed at each of $D$ steps,
which yields
$$W =O(Dn + Dm + D\rwh{\rho}C).$$

%
%

%
%
In {Erdős-Rényi graphs}, an edge exists with a probability $p$.
A balls-into-bins argument~\cite{Raab1998} may be applied to bound the maximum vertex degree, (each bin can be thought of as a vertex in the adjacency matrix and each ball an outgoing edge).
When $np=\Omega(\log(n))$, this analysis implies that the maximum degree of such a graph will be $\rwh{\rho}=O(np)$.
When $p$ is very small, the same type of argument leads to a bound of $\rwh{\rho}=O(\log(n))$.
%
Therefore, we ascertain
\begin{gather}
W = O(Dn + Dm + DC\log(n)) \label{eqn:W_ER}
\end{gather}

In power-law graphs, the probability that a vertex has degree $\rho$ is $f(\rho) = \alpha \rho^{-\beta}$.
To get a high-probability bound on the max degree of a power-law graph $\rwh{\rho}$, we 
consider the probability that a vertex has degree $\rho \geq \rwh{\rho}$,
\[P[\rho > \rwh{\rho}] =\alpha \sum_{x=\rwh{\rho}+1}^{n-1} x^{-\beta} \approx \alpha\int_{\rwh{\rho}}^{\infty} x^{-\beta}dx = \alpha\frac{\rwh{\rho}^{1-\beta}}{\beta-1}.\]
To ensure that with probability $1-1/\log(n)$ all vertices have degree less than $\rwh{\rho}$, we need
\[(1-P[\rho > \rwh{\rho}])^n \leq 1 - \frac{1}{\log(n)}.\]
This allows us to bound the maximum vertex degree with high probability (using Bernoulli's inequality),
\begin{align*}
P[\rho > \rwh{\rho}] &\geq 1- \Big(1-\frac{1}{\log(n)}\Big)^{1/n} \\
\rwh{\rho} &\geq \bigg(\frac{\beta-1}{\alpha} \bigg[1- \Big(1-\frac{1}{\log(n)}\Big)^{1/n}\bigg]\bigg)^\frac{1}{1-\beta} \\
&\geq \bigg(\frac{\beta-1}{\alpha} \frac{1}{n\log(n)}\bigg)^\frac{1}{1-\beta}=O\Big((\alpha n\log(n))^\frac{1}{\beta-1}\Big).
\end{align*}
Given this probabilistic bound on the maximum degree, we obtain a bound on the amount of work for power-law graphs,
\begin{gather}
W =O\Big(Dn+Dm+DC(\alpha n\log(n))^\frac{1}{\beta-1}\Big). \label{eqn:W_K}
\end{gather}

The analysis for other semirings differs in that more
instructions are required to derive $\mathbf{f}$ from
$\mathbf{x}$ in each iteration. These instructions take $O(n)$
of the total additional work to each term from the analysis for the tropical semiring.

%
%
%
%
%
%
%
%

\subsection{Reducing Storage Complexity with SlimSell}
\label{sec:slimsell}

\begin{figure*}[t]
    \centering
    \includegraphics[width=0.7\textwidth]{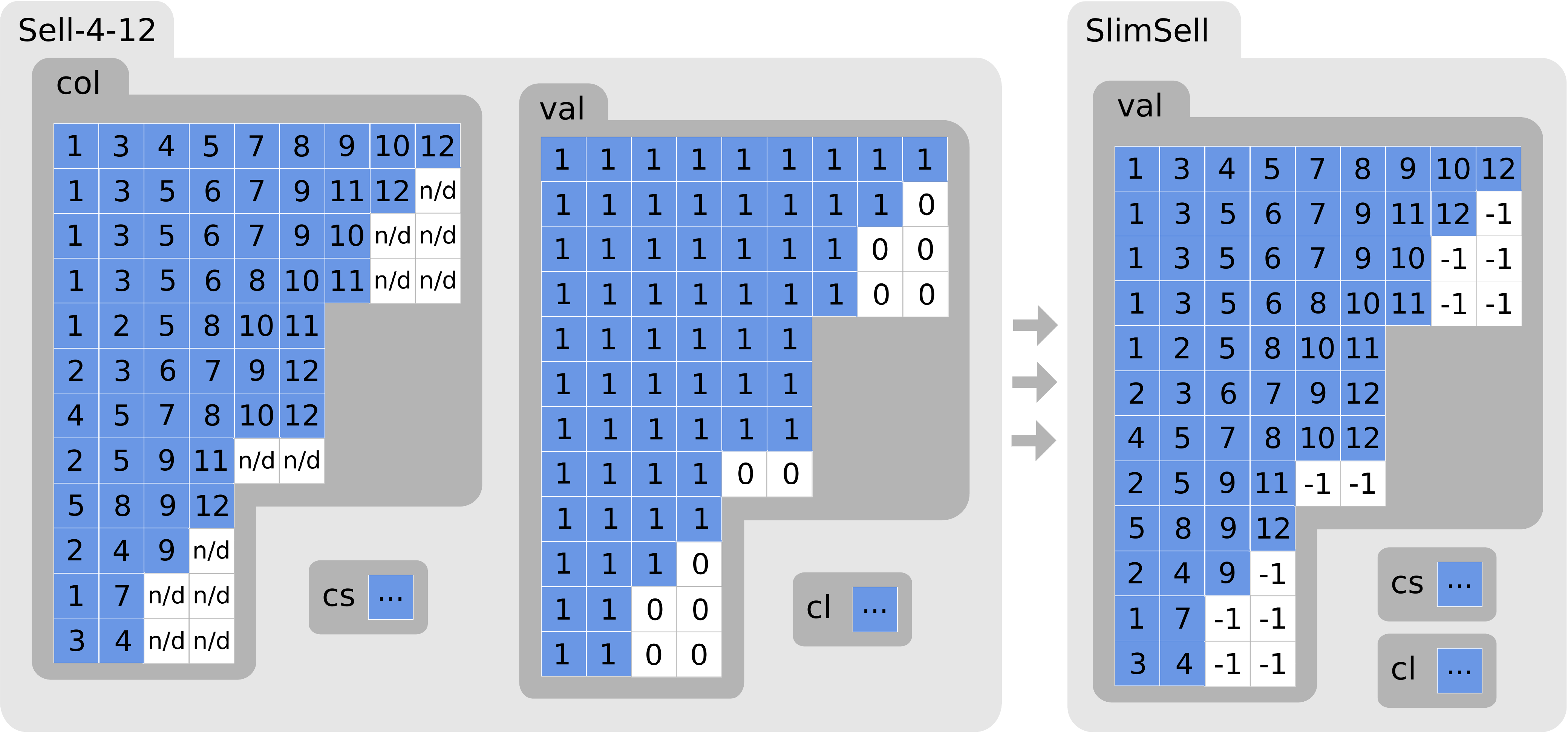}
    \caption{SlimSell: val can be inferred from col. $C = 4$, $\sigma = 12$.}
    \label{fig:slimsell}
\end{figure*}

We now build on Sell-$C$-$\sigma$ and introduce \emph{SlimSell}: a
representation for unweighted graphs that reduces the original
Sell-$C$-$\sigma$ size by a factor of up to $\frac{m+n}{2m+n}$, putting less
pressure on the memory subsystem.
Our key notion is that, for undirected graphs, entries in $\mathbf{A}$ only
indicate presence or absence of edges; this information can be contained in
$col$ without any need for $val$.
For this, each entry in $col$ in SlimSell contains either a usual column index of the
corresponding non-zero entry in $\mathbf{A}$ (as in Sell-$C$-$\sigma$) or a
special marker (e.g., $-1$) if the entry is not an edge but is only
present due to padding. Thus, an entry in $col$ different from the marker
indicates an edge and implies a $1$ in $val$. 
We illustrate SlimSell in Figure~\ref{fig:slimsell} and in
Listing~\ref{lst:min-plus-no-val}.
Instead of directly loading $val$ (like in Listing~\ref{lst:bfs-spmv-semirings},
line~\ref{line:load-val}), we first load the corresponding column indices from
$col$ (Listing~\ref{lst:min-plus-no-val}, line~\ref{line:no-val_load-cols}).
Then, a vectorized compare instruction provides us with a mask that determines
which column indices are $-1$ and which are non-negative entries corresponding
to edges. Finally, we derive the contents of $val$ (lines~\ref{line:no-val_vals-comp-start}-\ref{line:vals-final}).


\begin{lstlisting}[float=h!,label=lst:min-plus-no-val,caption=(\cref{sec:slimsell}) An example of using SlimSell; $C$ is $4$.]
// |\texttt{V}|ectors used by SlimSell:
V m_ones = [-1,...,-1], ones = [1,...,1], infs = [$\infty$,...,$\infty$];

for(int i = 0; i < $n$/$C$; i++) { // Iterate over each chunk.
  int index = $cs$[i]; // get offset to chunk i.
  V tmp = LOAD(&$f_{k-1}$[i*$C$]); // Load chunk i from frontier $f_{k-1}$.
  for(int j = 0; j < $cl$[i]; j++) { // Iterate over each column.
    V cols = LOAD(&$col$[index]); // Load entries from $col$. |\label{line:no-val_load-cols}|
    // Compute a mask that indicates which entry in cols is -1.
    V vals = CMP(cols, m_ones, EQ); |\label{line:no-val_vals-comp-start}|
    // Derive required entries from $val$.
    vals = BLEND(ones, infs, vals); |\label{line:vals-final}|
    // Create the rhs vector consisting of four $\mathbf{f}_{k-1}$ entries.
    V rhs = [$f_{k-1}$[$col$[index+3]], $f_{k-1}$[$col$[index+2]], 
             $f_{k-1}$[$col$[index+1]], $f_{k-1}$[$col$[index+0]]];
    tmp = MIN(ADD(rhs, vals), tmp); // Compute a new frontier.
    index += $C$;
  }
  STORE(&$f_k$[i*$C$], tmp); } // Set a new frontier.
\end{lstlisting}

\goal{Conclude and describe the consequences of our design}
SlimSell reduces data transfer by removing loads of
$val$. Yet, more computation is required
(lines~\ref{line:no-val_vals-comp-start}-\ref{line:vals-final}). We
show in~\cref{sec:slimsell_impact} that this does not entail significant
overheads and SlimSell offers more performance than Sell-$C$-$\sigma$.
%
%

\macbs{Storage Complexity}
SlimSell reduces the size of not only Sell-$C$-$\sigma$ but also CSR and AL;
see Table~\ref{tab:sizes} for a summary. 
%
%
Sell-$C$-$\sigma$ and SlimSell require respectively $4m + \frac{2n}{C} + P$ and
$2m + \frac{2n}{C} + P$ words; $P$ is the number of cells from padding and it
depends on a specific graph and its degree distribution.
%
%
SlimSell uses less storage than the corresponding AL if:

\footnotesize
\small
\begin{gather}
2m + \frac{2n}{C} + P < n+2m \Leftrightarrow P < n \left(1 - \frac{2}{C}\right)
\end{gather}
\normalsize
This translates to $P < \frac{3n}{4}$ for $C=8$ (e.g., in Xeon CPUs), $P < \frac{7n}{8}$
for $C=16$ (e.g., in KNLs), and $P < \frac{7n}{8}$ for $C=32$ (e.g., in Tesla GPUs).
%
%
%
We show in an empirical analysis that covers various families of graphs (\cref{sec:size}) that the impact of $P$ is
negligible for larger sorting scopes ($\sigma > \sqrt{n}$) 
and SlimSell
always uses less space than CSR and Sell-$C$-$\sigma$ and it is
comparable or more space-efficient than AL.
Finally, note that the size of $val$ in SlimSell and Sell-$C$-$\sigma$ ($= 2m + P$) is equal to the
amount of work $W$ of a single SpMV product. Thus, bounds~(\ref{eqn:W_ER}) and~(\ref{eqn:W_K})
equal storage complexities for Erdős-Rényi and power-law graph models. 

\begin{table}[h!]
\centering
\scriptsize
\sf
\begin{tabular}{l||l|l|l|l}
\toprule 
  Representation & Sell-$C$-$\sigma$ & CSR & AL & SlimSell \\ \midrule
  Size [cells]  & $4m + \frac{2n}{C} + P$ & $4m+n$ & $2m+n$ & $2m + \frac{2n}{C} + P$ \\ \bottomrule
\end{tabular}
\caption{(\cref{sec:slimsell}) Storage complexity of considered graph representations; $P$ is the number of padded memory cells.}
\label{tab:sizes}
\end{table}

\begin{tikzpicture}[remember picture,overlay,pin distance=0cm]
\draw[fill=black!10, opacity=1, inner sep=4pt, rounded corners=2pt]
  ([shift={(-0.25em,-4.5em)}]pic cs:sw-tr-1-s)
    rectangle
  ([shift={(23.5em,-6.6em)}]pic cs:sw-tr-1-e);
\fill ([shift={(20.4em,-5.25em)}]pic cs:sw-tr-1-e) node[rounded corners, text=white, fill=black!75, font=\tiny] {\macb{TROPICAL SEMIRING}};
\draw[fill=black!10, opacity=1, inner sep=4pt, rounded corners=2pt]
  ([shift={(-0.25em,-6.6em)}]pic cs:sw-bl-1-s)
    rectangle
  ([shift={(23.5em,-9.4em)}]pic cs:sw-bl-1-e);
\fill ([shift={(19.4em,-7.3em)}]pic cs:sw-bl-1-e) node[thin,rounded corners, text=white, fill=black!75, font=\tiny] {\macb{BOOLEAN \& REAL SEMIRING}};
\draw[fill=black!10, opacity=1, inner sep=4pt, rounded corners=2pt]
  ([shift={(-0.25em,-9.4em)}]pic cs:sw-bl-1-s)
    rectangle
  ([shift={(23.5em,-11.5em)}]pic cs:sw-bl-1-e);
\fill ([shift={(20.52em,-10.2em)}]pic cs:sw-bl-1-e) node[thin,rounded corners, text=white, fill=black!75, font=\tiny] {\macb{SEL-MAX SEMIRING}};
\end{tikzpicture}

\begin{lstlisting}[float=h,label=lst:slimwork,caption=(\cref{sec:slimwork-t}) SlimWork for various semirings; $C$ is $4$.]
// Frontier expansion: we start like in Listing |\ref{lst:min-plus-no-val}|.
for(int i = 0; i < $n$/$C$; i++) {
  int skip = 1; // Assume we skip the current chunk |\label{line:skip-start}|
  for(int j = 0; j < $C$; j++) {
|\tikzmark{sw-tr-s}|#ifdef USE_TROPICAL_SEMIRING |\label{line:slimwork-t-start}|
    if($f_{k-1}$[i*$C$+j] == $\infty$)   |\label{line:inf-check}|
      skip = 0; // If any distance is still $\infty$, go on.  |\label{line:slimwork-t-end}||\tikzmark{sw-tr-s}|
|\tikzmark{sw-tr-s}|#elif defined USE_BOOLEAN_SEMIRING $||$ |\label{line:slimwork-b-start}|
      defined USE_REAL_SEMIRING 
    if($g_{k-1}$[i*$C$+j] != 0) 
      skip = 0; // If any filter is still $\neq 0$, go on. |\label{line:slimwork-b-end}||\tikzmark{sw-tr-s}|
|\tikzmark{sw-tr-s}|#elif defined USE_SEL|\texttt{MAX}|_SEMIRING |\label{line:slimwork-s-start}|
    if($p_{k-1}$[i*$C$+j] != 0) 
      skip = 0; // If any parent is still $\neq 0$, go on. |\label{line:slimwork-s-end}||\tikzmark{sw-tr-s}|
#endif

  if(skip) // Swap the frontiers. 
    store(&$x_k$[i*$C$], load(&$x_{k-1}$[i*$C$])); continue; |\label{line:carry-result}|
  /* Now, continue with the MV product as in Listing |\ref{lst:bfs-spmv-semirings}|... */ }
\end{lstlisting}

\subsection{Reducing Work Amount with SlimWork}
\label{sec:slimwork}


SlimSell improves upon Sell-$C$-$\sigma$ by reducing it size.
We now introduce SlimWork (see Listing~\ref{lst:slimwork}), a scheme that
reduces the work amount in BFS-SpMV.  Intuitively, we skip chunks of
computation if they are associated with the final BFS values (distances or parent IDs) that do not change
in future iterations.
%
%
As our performance study shows (\cref{sec:performance_analysis}), this scheme
significantly accelerates BFS-SpMV. 
We now present how to use SlimWork with each semiring.

\subsubsection{Tropical Semiring}
\label{sec:slimwork-t}
SlimWork for the tropical semiring is illustrated in Listing~\ref{lst:slimwork}
(lines~\ref{line:slimwork-t-start}-\ref{line:slimwork-t-end}).  Consider
$\mathbf{f}_k = \mathbf{A'} \otimes \mathbf{f}_{k-1}$.  In the
label-setting BFS, ${f}_k^s$ can only differ from ${f}_{k-1}^s$ if
${f}_{k-1}^s$ is not the final distance.
By verifying whether ${f}_k^s = {f}_{k-1}^s$, we determine if ${f}_{k+1}^s$ has
to be computed or whether it can simply be carried over from $\mathbf{f}_{k}$
(line~\ref{line:carry-result}).
\subsubsection{Other Semirings} 

Real or boolean semirings require a vector
for the filtering term $\mathbf{g}_k = \overline{\sum_{l=0}^{k-1}
\mathbf{f}_l}$ (see~\cref{sec:r-semiring}) to verify if a given chunk of
$\mathbf{f}_k$ must be updated. The computation in a chunk is skipped if all
the rows in that chunk satisfy the condition that the corresponding entries of
$\mathbf{g}_k$ are $0$.
A similar criterion (based on the $\mathbf{p}_{k}$ vector) is used for the selmax
semiring; see Listing~\ref{lst:slimwork}.
%
%

\macbs{Work Complexity}
SlimWork adds $O(C)$ work per chunk
as it must decide whether to omit
chunks of computation.
However, in return it may reduce $W$ by $O(\rwh{\rho}_{i} C)$ for each skipped chunk~$i$; $\rwh{\rho}_i$
is the maximum degree in chunk~$i$.
\subsection{Improving Load Balance with SlimChunk}
\label{sec:slimchunk}

The final SlimChunk scheme aims at accelerating BFS-SpMV when executing on units with
large SIMD widths.
Specifically, consider a case with a large sorting scope (e.g., $\sigma = \sqrt{n}$).
The rows are sorted by the degree of the corresponding vertices. Thus, some
chunks (related to high-degree vertices) may entail much more computation than
the chunks related to low-degree vertices, resulting in load imbalance.  To
alleviate this, we split $\mathbf{A}$ not only horizontally into chunks, but we
also split the chunks vertically, in two dimensions. This still leads to the contiguous layout in
memory but also accelerates BFS-SpMV by dividing work more equally between threads, as we show
in~\cref{sec:performance_analysis}.

To conclude, SlimSell improves the design of BFS-SpMV along \emph{three dimensions}:
it reduces the size, lowers the amount of work, and improves the load balancing.

\begin{figure*}
    \centering
    \begin{subfigure}[t]{0.23 \textwidth}
        \centering
        \includegraphics[width=\textwidth]{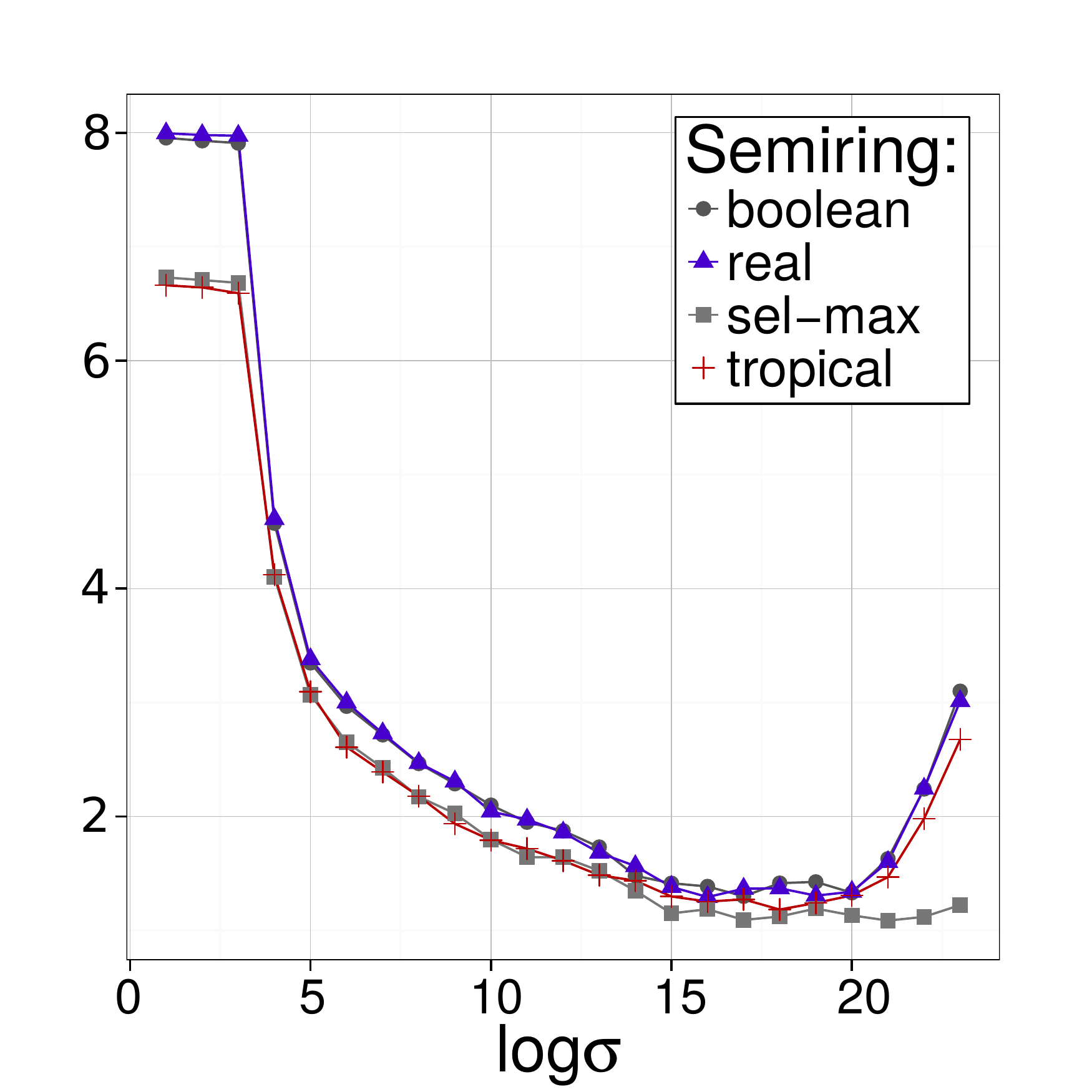}
        \caption{Kronecker, \texttt{DP}, \texttt{omp-s}.}
        \label{fig:cpu-K-dp-static}
    \end{subfigure}
    %
    %
     \begin{subfigure}[t]{0.23 \textwidth}
         \centering
         \includegraphics[width=\textwidth]{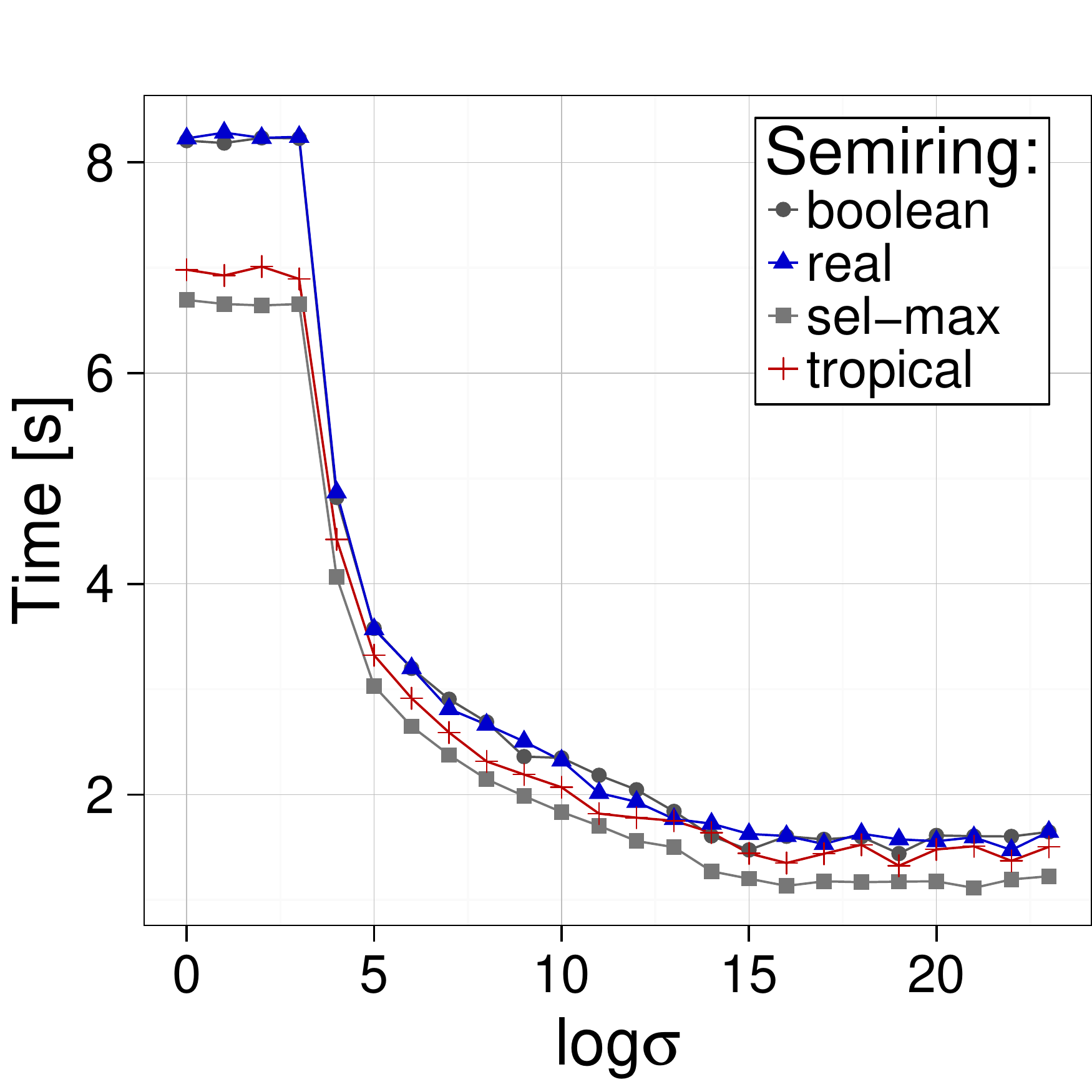}
         \caption{Kronecker, \texttt{No-DP}, \texttt{omp-d}.}
         \label{fig:cpu-K-dp-dynamic}
     \end{subfigure}
    %
    \begin{subfigure}[t]{0.23 \textwidth}
        \centering
        \includegraphics[width=\textwidth]{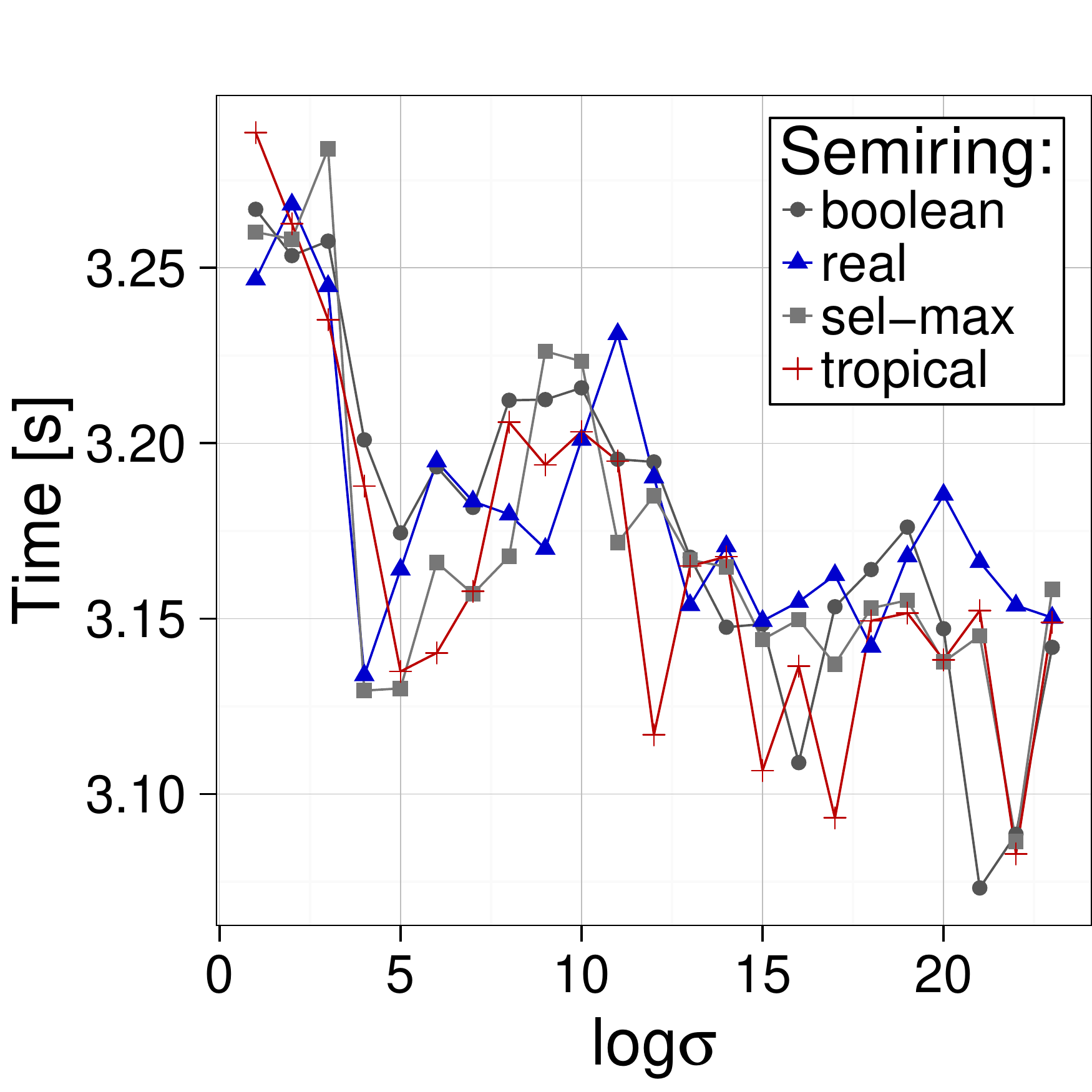}
        \caption{\texttt{ER}, \texttt{DP}, \texttt{omp-d}}
        \label{fig:cpu-E}
    \end{subfigure}
    %
    \begin{subfigure}[t]{0.23 \textwidth}
        \centering
        \includegraphics[width=\textwidth]{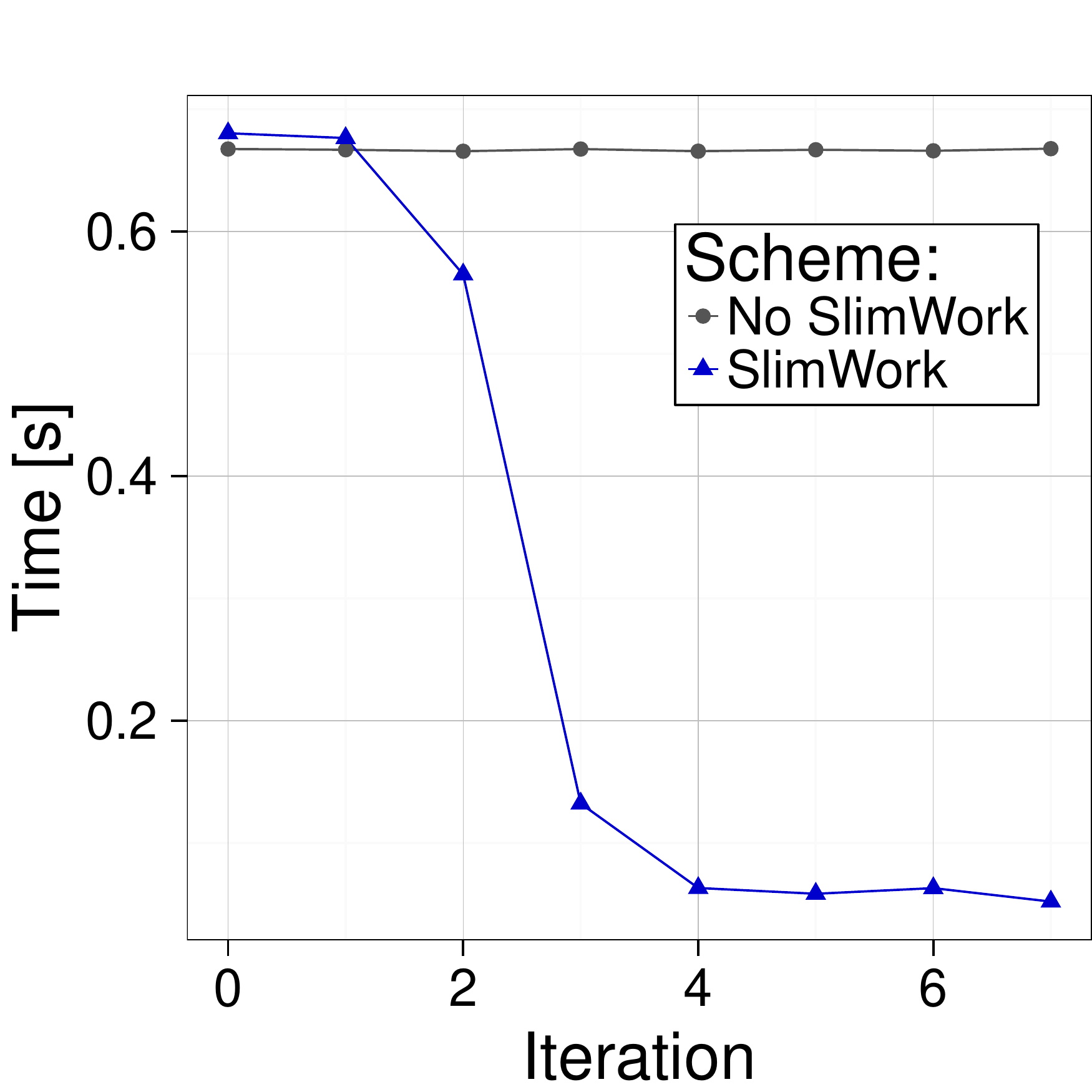}
        \caption{SlimWork.}
        \label{fig:cpu-slimwork}
    \end{subfigure}
    %
    \caption{(\cref{sec:cpu-analysis}) CPU analysis (Dora). The Kronecker graphs have
    $n=2^{23}, \bar{\rho}=16$; the \texttt{ER} graphs have $n = 2^{23}, p =
    2 \cdot 10^{-6}, \bar{\rho} \approx 16$.} \label{fig:cpu}
\end{figure*}

\section{Performance Analysis and Evaluation}
\label{sec:performance_analysis}


\maciej{More comparison targets I found: https://github.com/gunrock/gunrock (gunrock library),
the code I sent: (https://github.com/iHeartGraph/iBFS), this GPU g500 code we know,
then there is this Cusha library: http://farkhor.github.io/CuSha/, there seems to be
also some opensource implemention that accompanies the Merill paper. 
Please try to find other targets as well, so that we know exhaustively what
is out there. If we get rejected but resumit with all these comparisons,
or if we have a camera-ready IPDPS paper where we show we beat these targets
and really run them all, we will have a high chance of a best paper award}

We now illustrate that SlimSell secures high performance for BFS-SpMV.
%
%
We also analyze the 
differences between BFS-SpMV based on various semirings and representations.
%


%

\macbs{Selection of Benchmarks and Parameters}
We consider all four BFS-SpMV semiring variants. 
%
%
Next, performance and storage improvements from SlimSell, 
SlimWork, and SlimChunk are evaluated. Both strong- and weak-scaling is considered.
%
%
We also analyze the $DP$ transformation overheads (\texttt{DP},
\texttt{No-DP}).
To evaluate the impact from sorting, we vary $\sigma \in $ [$1$,$n$] (for
simplicity $\sigma$ is a multiple of $C$). We verify the BFS
performance in each iteration for a fine-grained analysis. Next, static and
dynamic OpenMP scheduling is incorporated (\texttt{omp-s}, \texttt{omp-d}).
Three classes of graphs are considered: synthetic power-law 
Kronecker~\cite{leskovec2010kronecker} and Erdős-Rényi
(\texttt{ER})~\cite{erdHos1976evolution} graphs such that $n \in \{2^{20}, ...,
2^{28}\}$ and $\overline{\rho} \in \{2^1, ..., 2^{10}\}$; we also use real-world
graphs (\texttt{RW}); see Table~\ref{tab:graphs}.
We present and discuss in detail a representative 
subset of results.

\begin{table}[h]
\centering
\scriptsize
\sf
\begin{tabular}{@{}l|lrrrr@{}}
\toprule
\textbf{Type}                                                                           & \multicolumn{1}{l}{\textbf{ID}} & \multicolumn{1}{r}{\textbf{$n$}} & \multicolumn{1}{r}{\textbf{$m$}} & \multicolumn{1}{r}{\textbf{$\bar{\rho}$}} & \multicolumn{1}{r}{\textbf{$D$}} \\ \midrule
\multirow{1}{*}{Kronecker graphs}          & K                                    & 1M-268M                          & 2M-536M          &   2-1014  & 19-33           \\ \midrule
\multirow{3}{*}{Social networks}
                                                                                        & orc                                 & 3.07M                          & 117M          &      39      & 9    \\
                                                                                        & pok                                 & 1.63M                          & 30.6M          &      18.75   &  11     \\ 
                                                                                        & epi                  & 75k                            & 508k             &     6.7  & 15        \\ \midrule
\multirow{1}{*}{Ground-truth~\cite{yang2015defining} community}
                                                                                        & ljn                                 & 3.99M                          & 34.6M          &    8.67   & 17       \\ \midrule
\multirow{4}{*}{\begin{tabular}[c]{@{}c@{}}Web graphs\end{tabular}}             & brk                                     & 685k                           & 7.60M           &  11.09        & 514     \\
                                                                                         & gog                                & 875k                           & 5.1M                &     5.82   & 21 \\
                                                                                        & sta                                     & 281k                           & 2.31M           &   8.2      & 46      \\ 
                                                                                        & ndm                                    & 325k                           & 1.49M           &    4.59    & 674       \\ \midrule
\multirow{1}{*}{Purchase network} 
                                                                                        & amz                                   & 262k                           & 1.23M           &   4.71  & 32    \\ \midrule
\multirow{1}{*}{Road network}          & rca                                   & 1.96M                          & 2.76M          &   1.4  & 849           \\ \bottomrule
\end{tabular}

\caption{The analyzed graphs with skewed degree distributions.}
\label{tab:graphs}
\end{table}

\macbs{Comparison Targets (Performance)}
We compare BFS-SpMV equipped with SlimSell to the work-efficient highly-optimized
OpenMP BFS Graph500 code~\cite{murphy2010introducing}
(\texttt{Trad-BFS}). This baseline applies several
optimizations; among others it reduces the amount of fine-grained
synchronization by checking if the vertex was visited before executing an
atomic.
As \texttt{Trad-BFS} is work-optimal, it also represents the work-optimal BFS
based on SpMSpV.
We execute \texttt{Trad-BFS} on both KNLs and CPUs to secure fair comparison and to
illustrate \emph{one of the points of our work: vectorization- and SIMD-friendly SlimSell
enables BFS-SpMV to be comparable to work-efficient BFS optimized for CPUs.}

\macbs{Comparison Targets (Storage)}
In the storage analysis, we compare SlimSell to CSR, AL, and Sell-$C$-$\sigma$.

%

\macbs{Experimental Setup and Architectures}
We conduct the analysis on several CPUs, GPUs, and manycores, with
the \emph{total of seven evaluated systems}:
\begin{description}[leftmargin=0.5em]
\item[Dora (CPU)] One node of the CSCS Piz Dora supercomputer
contains two Intel Xeons E5-2695 v4 @ 2.10 GHz. Each CPU has 18 cores, 18x 256
KB L2 and 45 MB L3 cache. Each nodes has 64 GB of RAM. We use gcc-5.3. 
\item[Tesla (GPU)] We use NVIDIA Tesla K80 GPUs from the CSCS Greina HPC
cluster (covering high-performance GPUs).  Any CPU code is compiled using
gcc-5.3, the CUDA code uses cudatoolkit-6.5. 
%
%
\item[KNL (manycore)] We use Intel Xeon Phi KNLs 7210 (also from
Greina). 
Each KNL has 64 4-way 1.3GHz multi-threaded cores with the total of
32 MB L2 and 66 GB of RAM. 
%
%
We use icc 16.0.1.150. 
\item[Others] Other systems include a Tesla K20X from the CSCS Piz Daint supercomputer, a
Trivium server with an Intel Haswell CPU and an NVIDIA GTX 670 that represent
commodity machines, and a Xeon E5-1620 @3.50GHz from CSCS Greina
to cover very low-latency systems.
\end{description}

\begin{figure*}
    \centering
    \begin{subfigure}[t]{0.2 \textwidth}
        \centering
        \includegraphics[width=\textwidth]{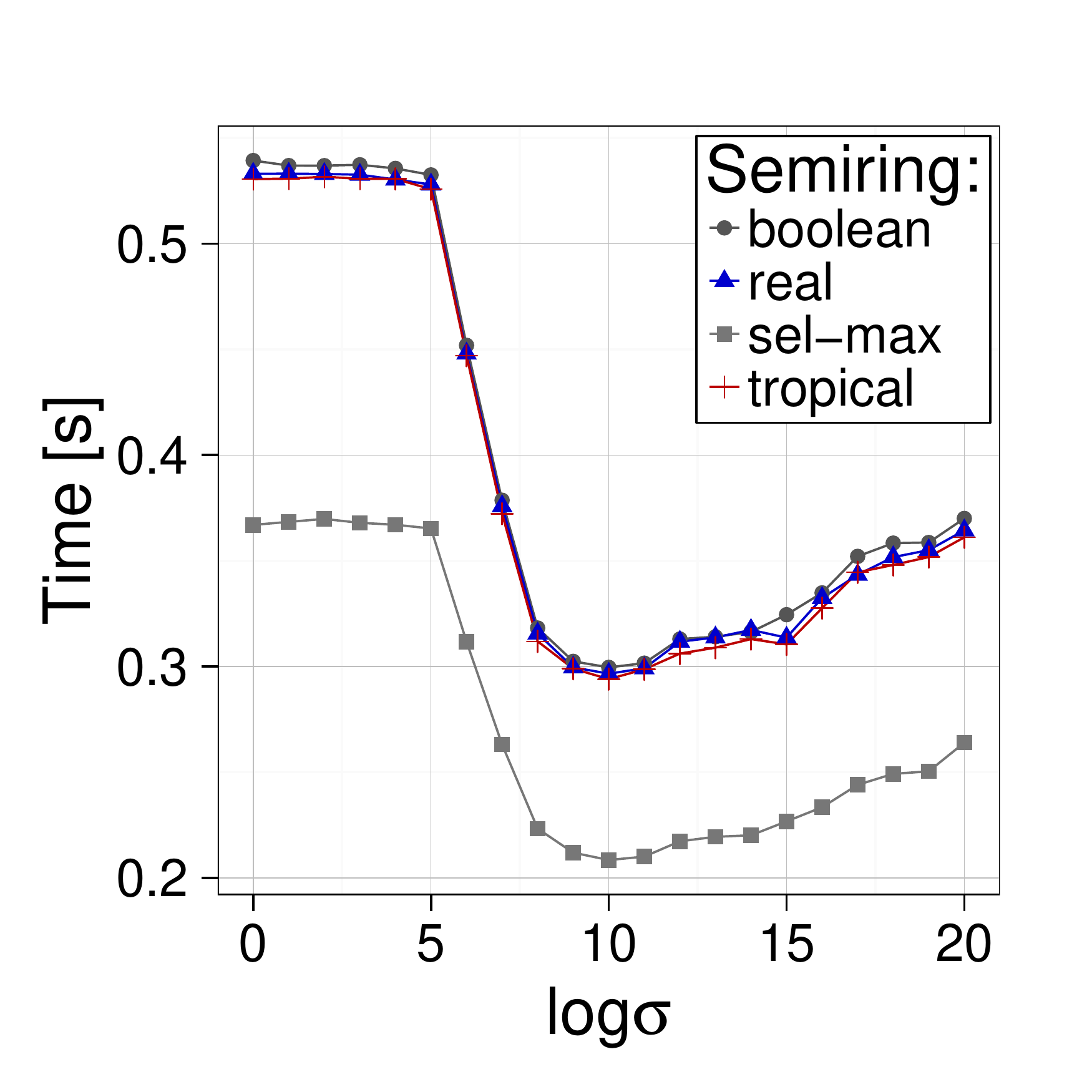}
        \caption{Kronecker, \texttt{DP}.}
        \label{fig:gpu_sigma}
    \end{subfigure}
    \hspace{-1em}
    \begin{subfigure}[t]{0.2 \textwidth}
        \centering
        \includegraphics[width=\textwidth]{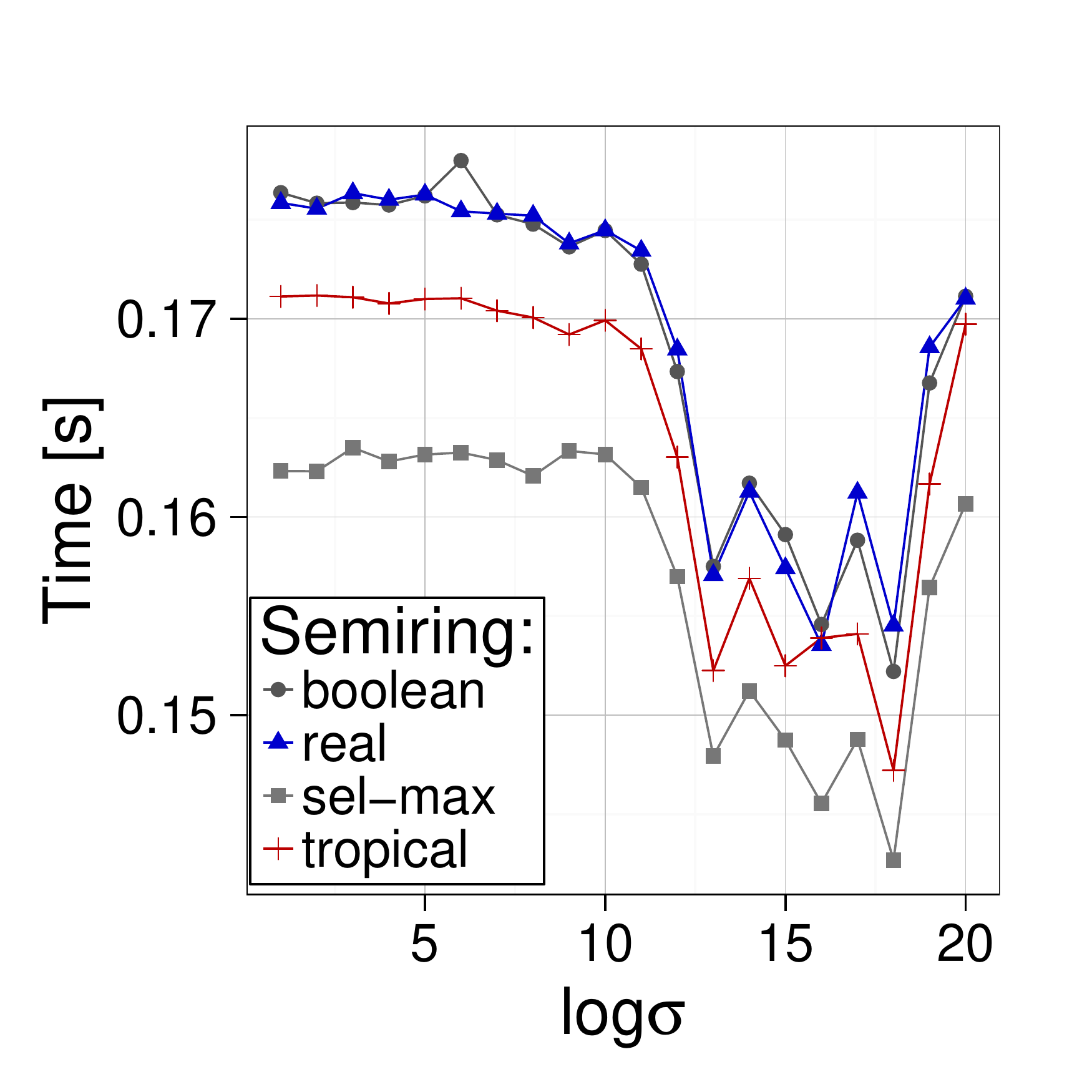}
        \caption{\texttt{ER}, \texttt{DP}.}
        \label{fig:gpu_sigma_erdos}
    \end{subfigure}
    \hspace{-1em}
    \begin{subfigure}[t]{0.2 \textwidth}
        \centering
        \includegraphics[width=\textwidth]{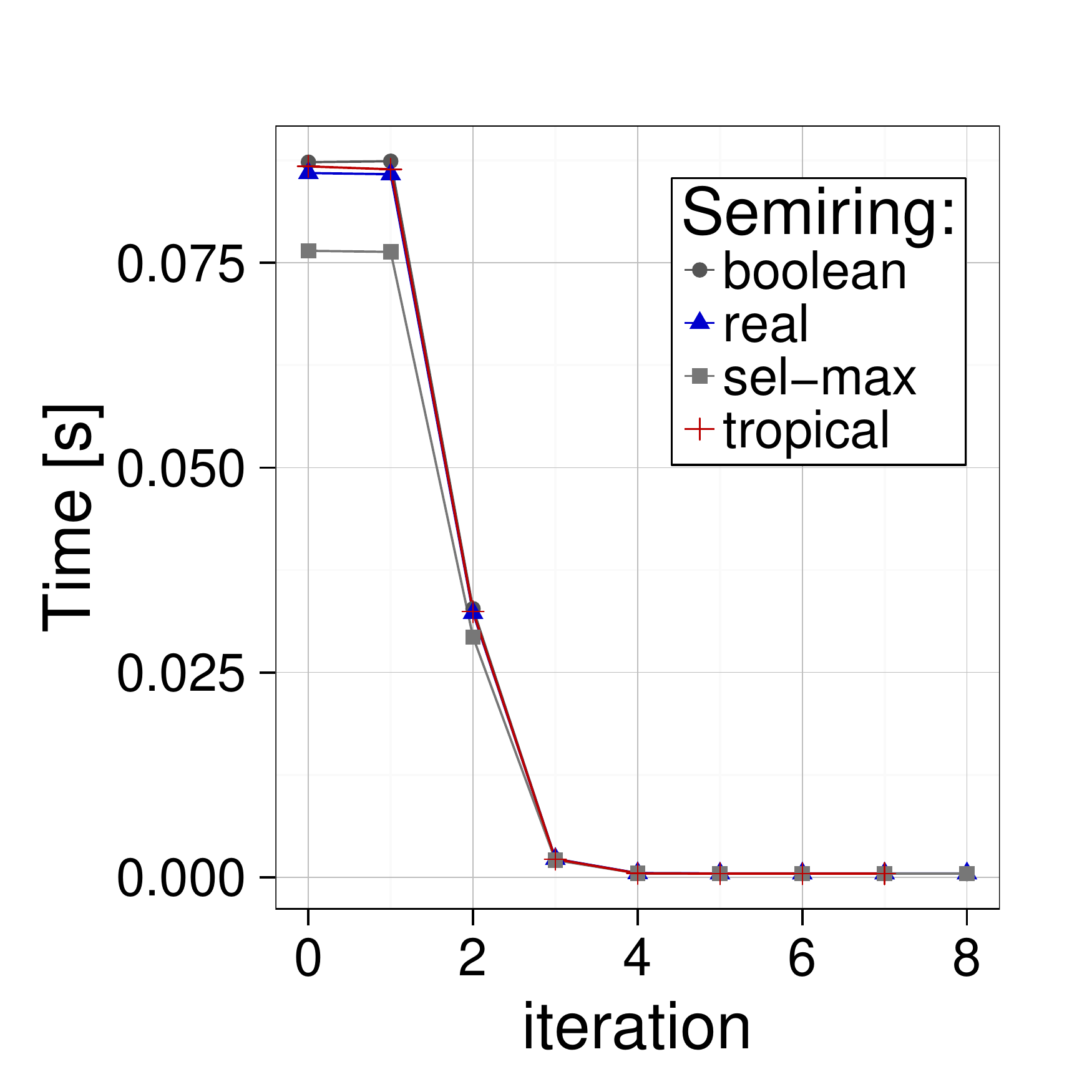}
        \caption{Kronecker, \texttt{DP}, $\sigma = 2^{10}$.}
        \label{fig:gpu_per-it_kron}
    \end{subfigure}
    \hspace{-1em}
    \begin{subfigure}[t]{0.2 \textwidth}
        \centering
        \includegraphics[width=\textwidth]{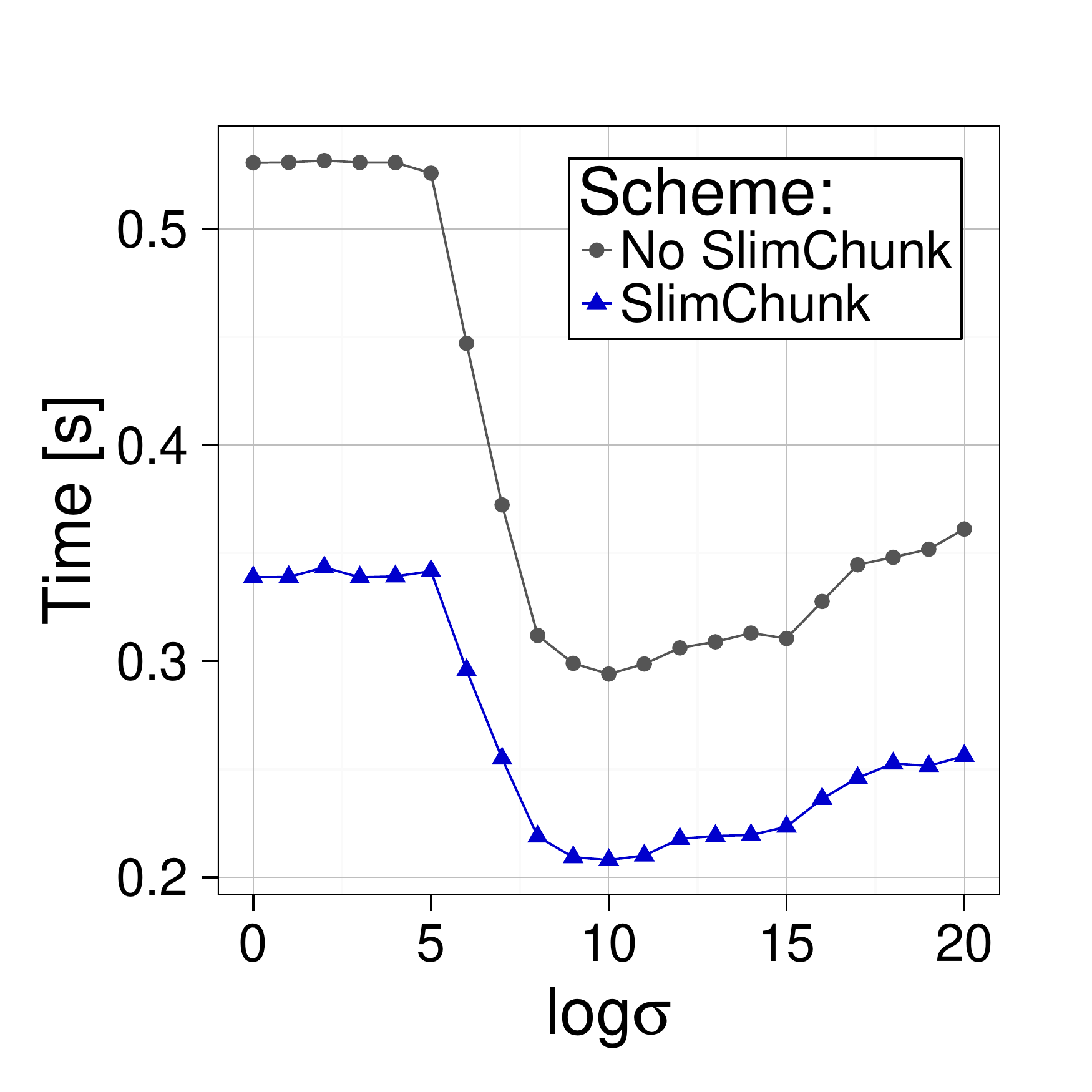}
        \caption{Kronecker, \texttt{DP}.}
        \label{fig:gpu_sigma_improved}
    \end{subfigure}
    \hspace{-1em}
    \begin{subfigure}[t]{0.2 \textwidth}
        \centering
        \includegraphics[width=\textwidth]{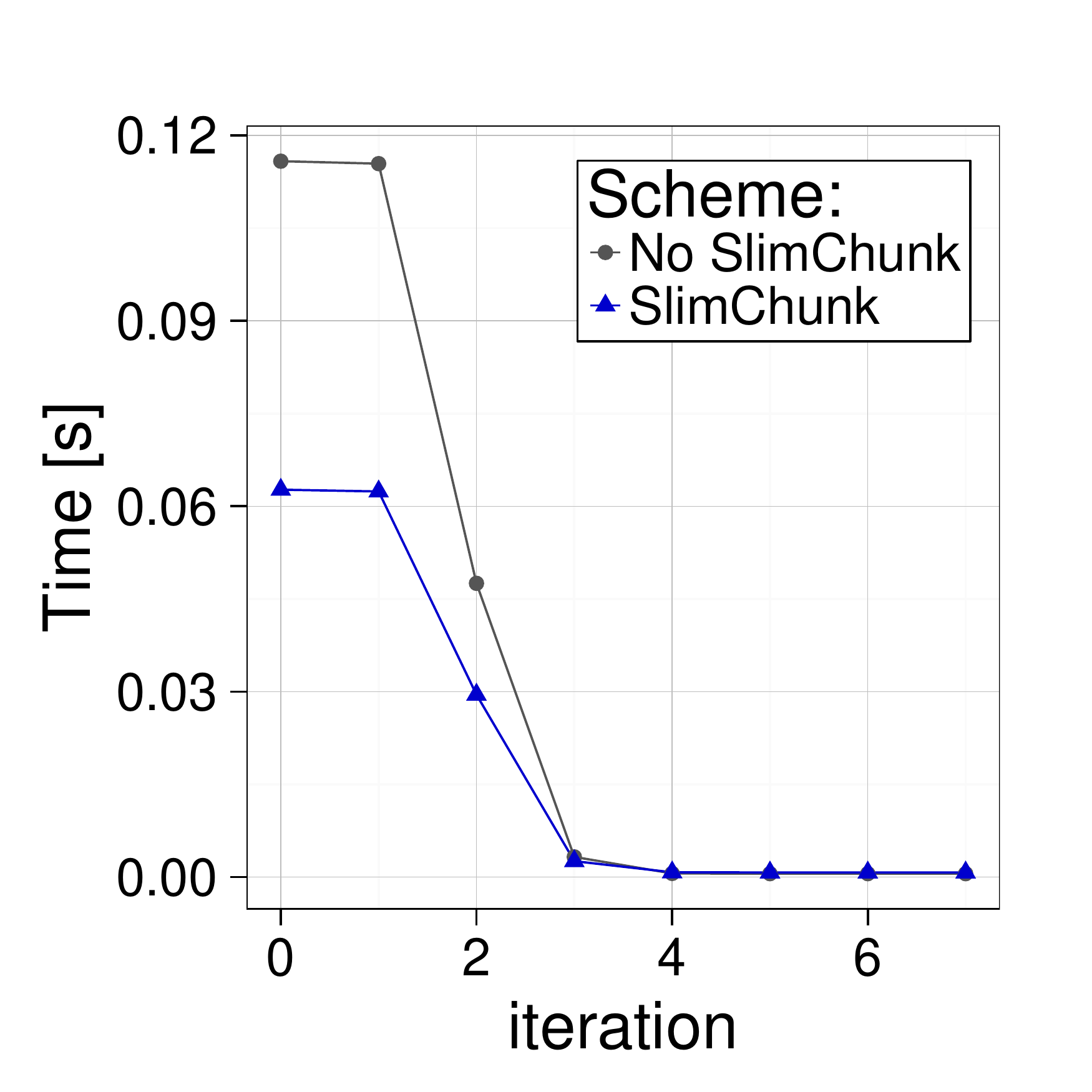}
        \caption{Kronecker, \texttt{DP}, $\sigma = 2^{10}$.}
        \label{fig:gpu_per-it_improved}
    \end{subfigure}
    %
    \caption{(\cref{sec:gpu-analysis}) GPU analysis (Tesla K80). The Kronecker graphs have
    $n=2^{20}, \bar{\rho}=16$; the \texttt{ER} graphs have $n = 2^{18}, p =
    1.5 \cdot 5 \cdot 10^{-5}, \bar{\rho} \approx 16$.} \label{fig:gpu}
\end{figure*}


\subsection{CPU Analysis (Xeon E5-2695 v4)}
\label{sec:cpu-analysis}

We start with the CPU analysis; see Figure~\ref{fig:cpu}.
We fix $C$ to match architectural constraints. For example, choosing 32-bit
integers to represent vertex identifiers on a CPU yields a SIMD width of 8
due to 256-bit wide AVX registers. 
We use OpenMP for intra-node parallelization.

\subsubsection{Impact from Sorting ($\sigma$)}
%
We first investigate the impact of $\sigma$ for various BFS-SpMV scenarios
(Figures~\ref{fig:cpu-K-dp-static}-\ref{fig:cpu-E}).  Performance strongly
depends on $\sigma$ as significantly different vertex degrees in one $C$ entail
wasted work due to padding. Thus, large $\sigma$ is preferable.
Next, $\sigma < C$ does not reduce padding and work but only reorders rows
within a chunk.  Thus, the performance does not increase while $\log{\sigma}
\le 3$.
Moreover, in Figure~\ref{fig:cpu-K-dp-static}, an overhead due to the $DP$
transformation is visible as $\sigma$ goes towards $n$, except for the sel-max
semiring.
%
%
This is because in static OpenMP scheduling all threads process similar counts
of chunks. Then, for a very large $\sigma$, the first chunk contains all of the
longest rows and consequently the corresponding thread performs the majority of
work, causing imbalance. 
Dynamic OpenMP scheduling adds $\approx$1-2\% of relative overhead but
alleviates the problem; see Figure~\ref{fig:cpu-K-dp-dynamic}.

\subsubsection{Differences between Semirings}

The inner-most loop that processes chunks is the same for each
semiring except for the used vector instructions (cf.\ Listing~\ref{lst:bfs-spmv-semirings},
lines~\ref{line:mv-inner-start}-\ref{line:mv-inner-end}). As Figure~\ref{fig:cpu} shows,
performance differences are negligible ($\approx$1-2\%), confirming
the intuition that BFS is memory-bound.
However, semirings also differ in the post-processing that derives $\mathbf{f}_k$ from
$\mathbf{x}_k$ (lines~\ref{line:mv-post-start}-\ref{line:mv-post-end}).
The tropical semiring contains no additional work. Contrarily, the real and boolean semirings
update the filter $\mathbf{g}_k$ (six instructions and two stores) while the sel-max
semiring processes the $\mathbf{p}$ vector (four instructions and two stores).
Still, the only major performance difference comes with the DP in sel-max semiring.

\subsubsection{Advantages of SlimSell}
\label{sec:slimsell_impact}

We next illustrate advantages of SlimSell over Sell-$C$-$\sigma$
(Table~\ref{tab:slimsell}).  Here, $\sigma$ determines how much bandwidth is
wasted on unused loads. Larger $\sigma$ results in lower impact from SlimSell
as the bandwidth is used more efficiently due to sorting.

\begin{table}[h!]
\centering
\scriptsize
\sf
\begin{tabular}{l|llll}
\toprule
         $\sigma$ & Boolean & Real & Tropical & Sel-max \\ \midrule
         $2^4$  &$1.17$ & $1.17$ & $1.21$ & $1.18$ \\ 
         $2^{18}$ &$1.00$ & $1.04$ & $1.04$ & $1.03$ \\ \bottomrule
    \end{tabular}
\caption{Speedup of SlimSell over Sell-$C$-$\sigma$ in Kronecker graphs ($n = 2^{24}, \overline{\rho} = 16$).}
\label{tab:slimsell}
\end{table}

\subsubsection{Advantages of SlimWork}
\label{sec:slimwork_impact}

We present SlimWork advantages in Figure~\ref{fig:cpu-slimwork}.
We compare BFS-SpMV variants with and without SlimWork for several $\sigma$s
(for clarity, we plot only the case with $\sigma = n$).
The larger $\sigma$, the faster work amount decreases with iterations.
This is because larger $\sigma$s make it more probable that
the majority of the largest chunks (associated with high-degree vertices)
are processed in early iterations.
Next, SlimWork skips
chunks as soon as all the associated vertices have been visited. Thus, as more
vertices are visited, the number of chunks left to compute decreases and the last
few iterations entail only
little work.
Contrarily, early iterations gain no speedup; they instead pay a
small overhead for checking the skipping criteria. 
Still, the overhead is rapidly outweighed by saving expensive chunk computations.
Overall, SlimWork accelerates the baseline SlimSell by a large factor, for example
$\approx$129\% for sel-max for a Kronecker graph with $n = 2^{24}, \bar{\rho} = 16$.
Finally, in ``No SlimWork''
(corresponding to Sell-$C$-$\sigma$ augmented only with the SlimSell storage reductions),
there is no performance improvement after the first iteration.

\begin{figure*}
    \centering
    \begin{subfigure}[t]{0.49 \textwidth}
        \centering
        \includegraphics[width=\textwidth]{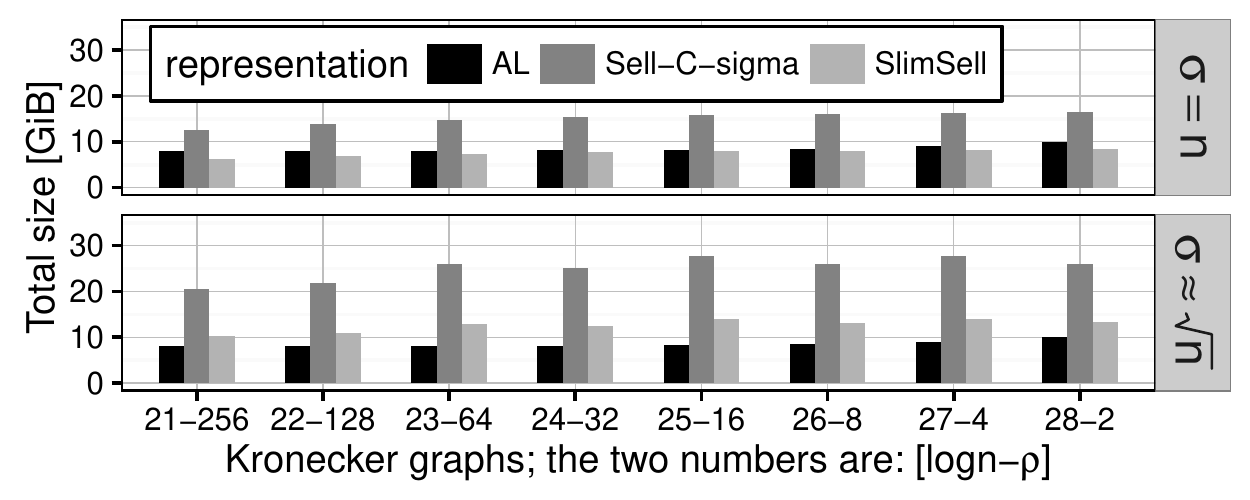}
        \caption{Kronecker graphs (larger sorting scope).}
        \label{fig:size-kronecker-more-sort}
    \end{subfigure}
    \begin{subfigure}[t]{0.49 \textwidth}
        \centering
        \includegraphics[width=\textwidth]{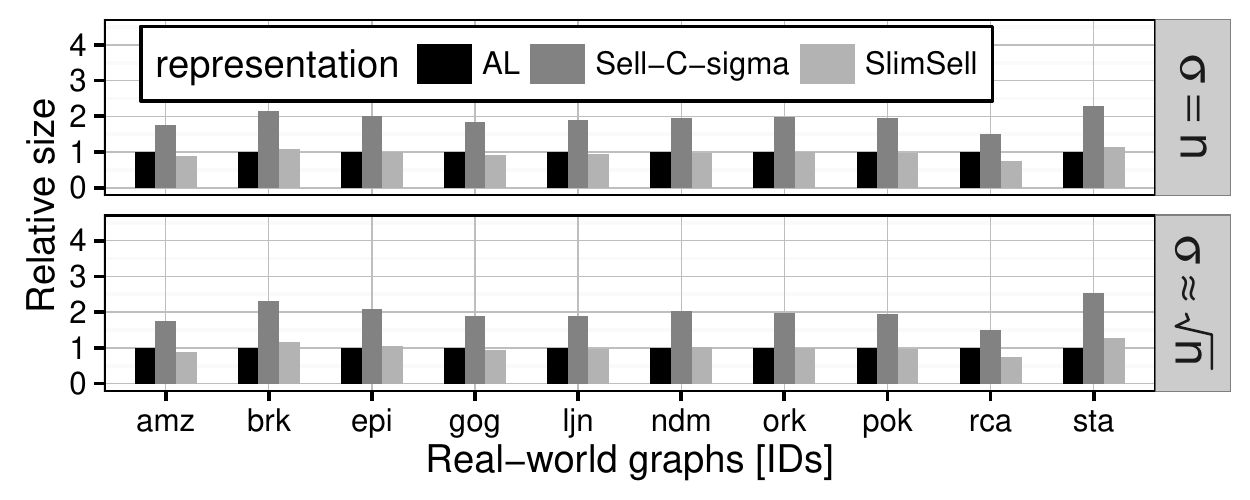}
        \caption{Real-world graphs (larger sorting scope).}
        \label{fig:size-rw-more-sort}
    \end{subfigure}\\
    \begin{subfigure}[t]{0.49 \textwidth}
        \centering
        \includegraphics[width=\textwidth]{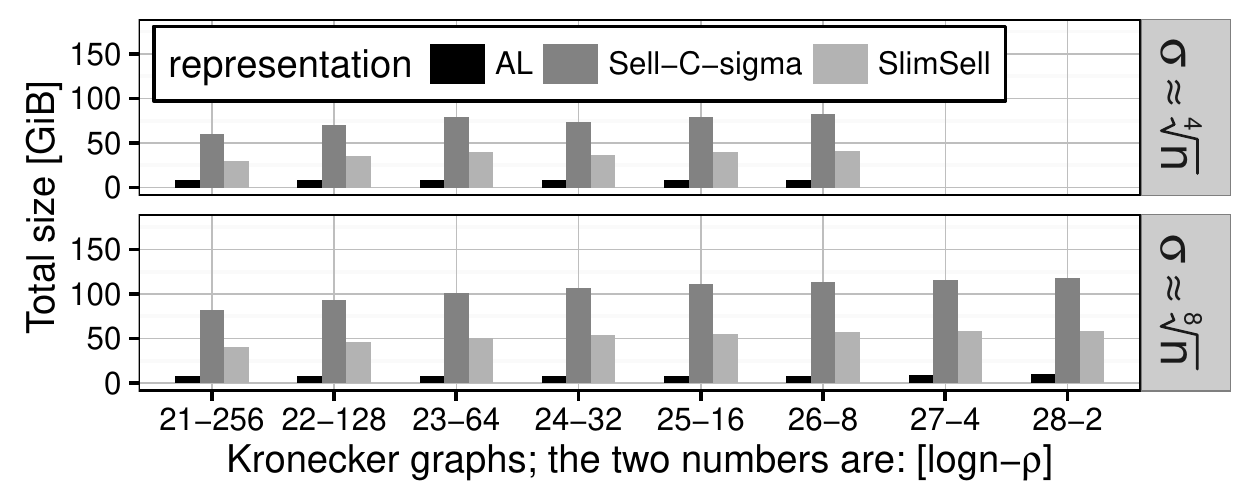}
        \caption{Kronecker graphs (smaller sorting scope).}
        \label{fig:size-kronecker-more-sort}
    \end{subfigure}
    \begin{subfigure}[t]{0.49 \textwidth}
        \centering
        \includegraphics[width=\textwidth]{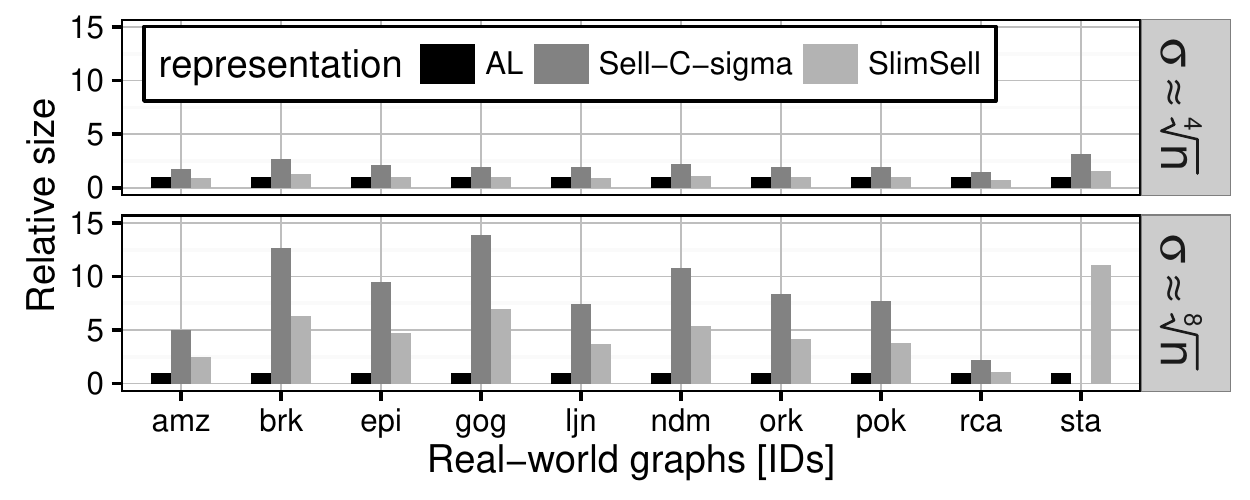}
        \caption{Real-world graphs (smaller sorting scope).}
        \label{fig:size-rw-more-sort}
    \end{subfigure}
    %
    \caption{(\cref{sec:size}) Size analysis for various $n$, $\bar{\rho}$, and $\sigma$; the used Kronecker graphs fill the whole memory; $C$ is 8.} \label{fig:size}
\end{figure*}

\subsubsection{Various Graph Families}
\label{ssec:graph_family}

%
We now discuss Erdős-Rényi and real-world inputs.
The degrees of the former follow the uniform instead of the power law
distribution. The impact from large $\sigma$ is therefore less evident; see
Figure~\ref{fig:cpu-E}. The fine-grained analysis of each BFS step
illustrates that the decrease in time only starts during the final iterations.
Next, we consider real-world graphs. Some, like \texttt{amz} or \texttt{rca},
have high $D$ and low $\bar{\rho}$ ($3.4$ for \texttt{amz} and $1.4$ for
\texttt{rca}). This leads to small or no improvement from SlimWork, regardless
of $\sigma$. 
%
Others behave similarly to
Kronecker graphs as their $\bar{\rho}$ is large enough for SlimWork to ensure
speedups.
\subsection{GPU Analysis (Tesla K80)}
\label{sec:gpu-analysis}

We present a representative subset of results in Figure~\ref{fig:gpu}.
Our insights are similar to those of the CPU analysis.
The main difference is that wider SIMD units in GPUs secure higher
speedups.
First, Figures~\ref{fig:gpu_sigma} and~\ref{fig:gpu_sigma_erdos} depict the total execution time for varying $\sigma$
for Kronecker and \texttt{ER} graphs.
Similarly to CPUs, the performance
does not improve up to a certain $\sigma$ equal to $C$. 
Second, sel-max performs better than others as it does not need $DP$. 
Moreover, the performance for higher $\sigma$ 
is because the sorting of the rows in $\mathbf{A}$ leads to
large load imbalance (e.g., see Figure~\ref{fig:gpu_sigma_erdos}).
Then, the per iteration performance follows patterns similar to CPUs
as illustrated in Figure~\ref{fig:gpu_per-it_kron}.
Finally, we show that SlimChunk brings expected significant advantages, for example
$\approx$50\% in the first two iterations in
%
Figure~\ref{fig:gpu_per-it_improved}).

\begin{figure}
    \centering
    \begin{subfigure}[t]{0.235 \textwidth}
        \centering
        \includegraphics[width=\textwidth]{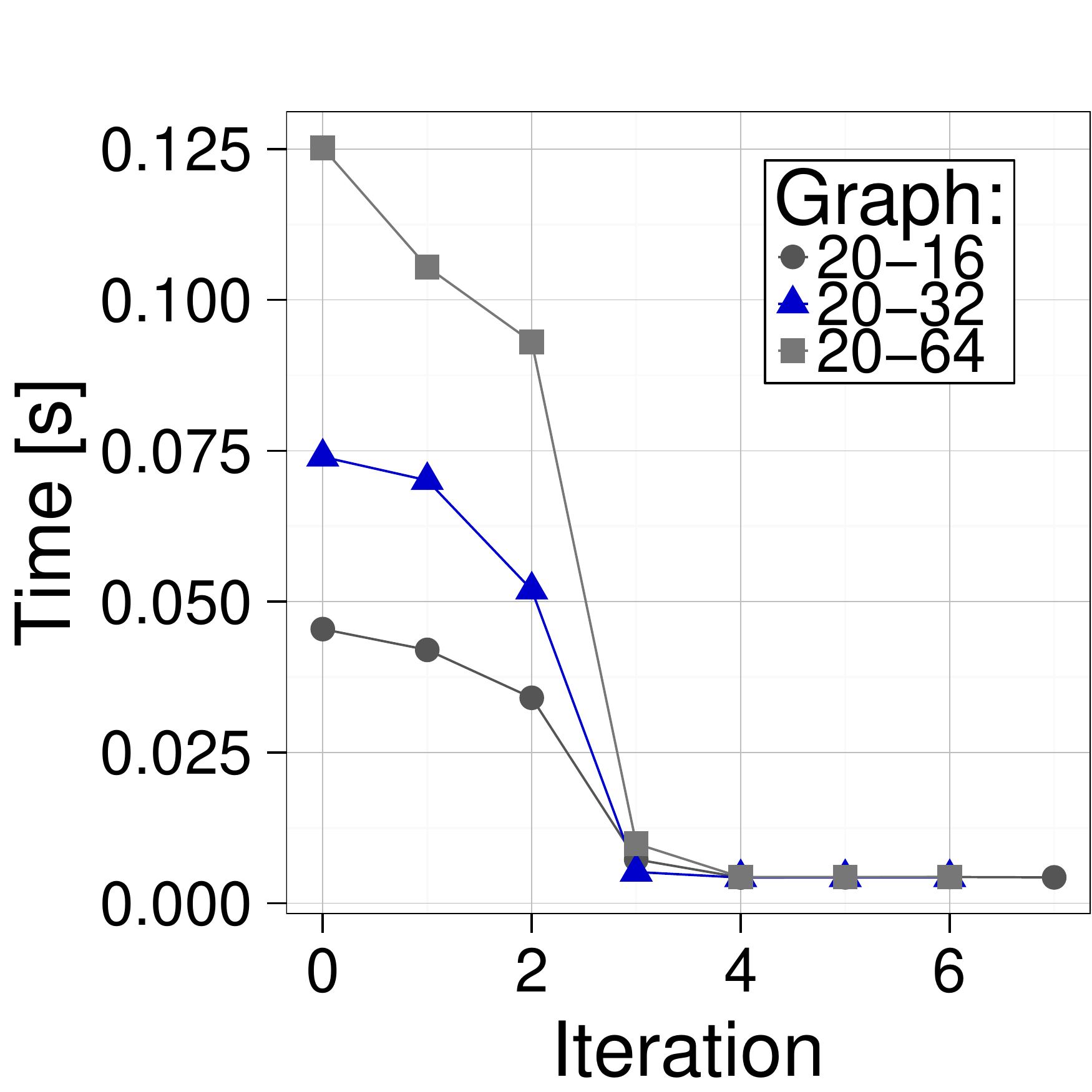}
        \caption{$n = 2^{20}$.}
    \end{subfigure}
    \hspace{-0.5em}
    \begin{subfigure}[t]{0.235 \textwidth}
        \centering
        \includegraphics[width=\textwidth]{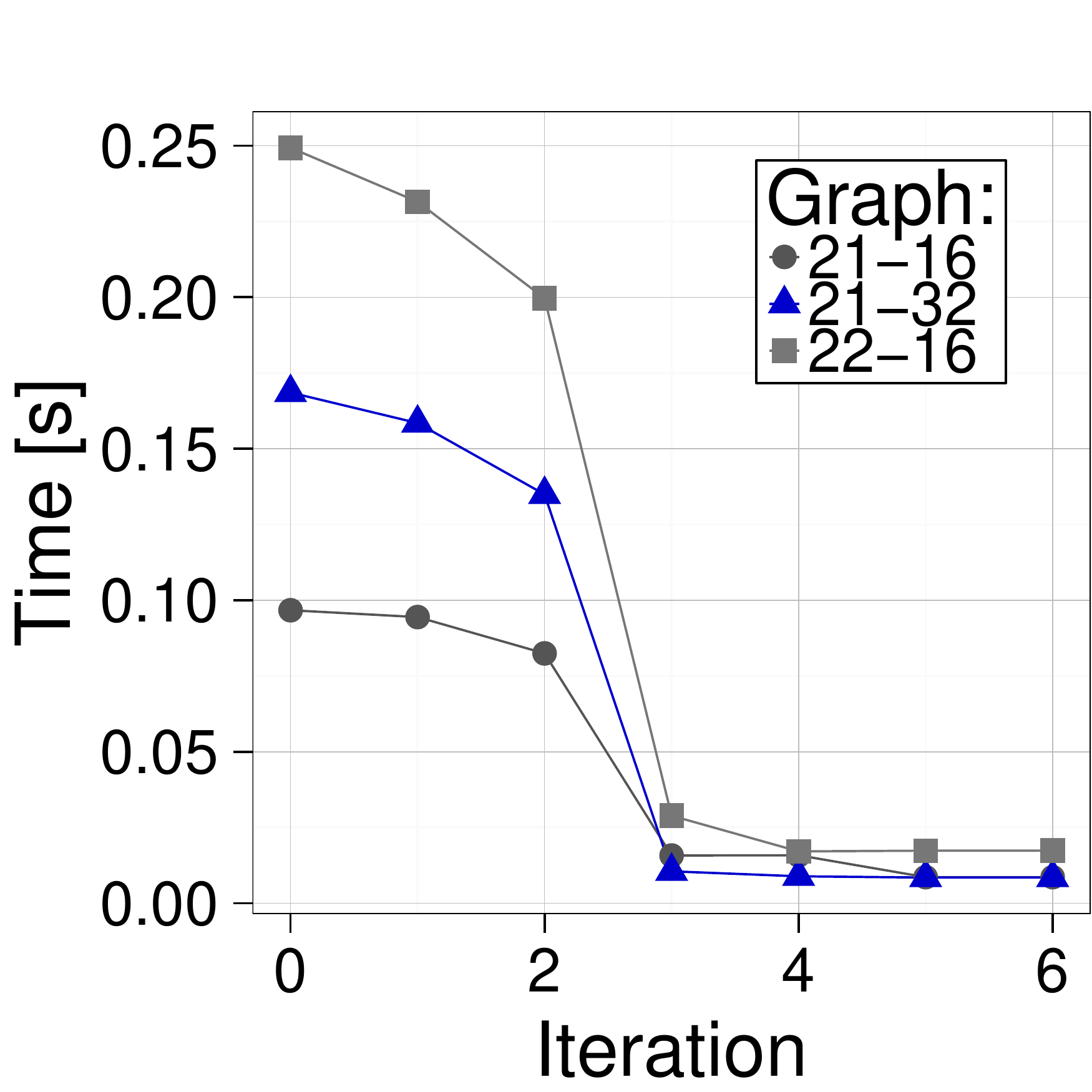}
        \caption{$n \in \{2^{21}, 2^{22}\}$.}
    \end{subfigure}
    %
    \caption{(\cref{sec:knl}, KNL analysis) A fine-grained analysis (using the tropical semiring) on
    Kronecker graphs; two numbers for each graph are $\log
    n$-$\overline{\rho}$.} \label{fig:knl}
\end{figure}

\begin{figure*}[t]
    \centering
    \begin{subfigure}[t]{0.2 \textwidth}
        \centering
        \includegraphics[width=\textwidth]{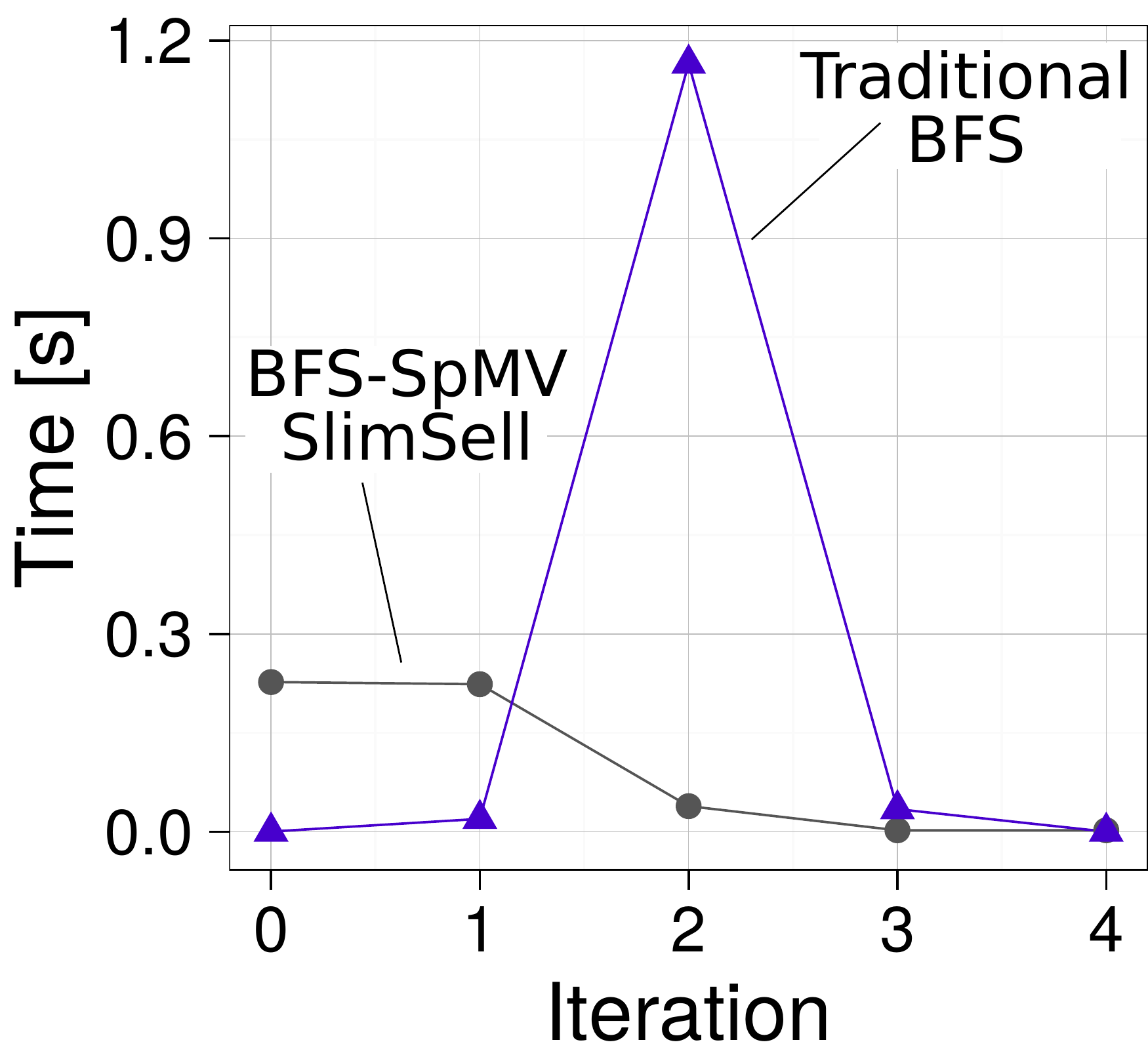}
        \caption{$n=2^{19}, \overline{\rho} = 1024$}
    \end{subfigure}
    \begin{subfigure}[t]{0.2\textwidth}
        \centering
        \includegraphics[width=\textwidth]{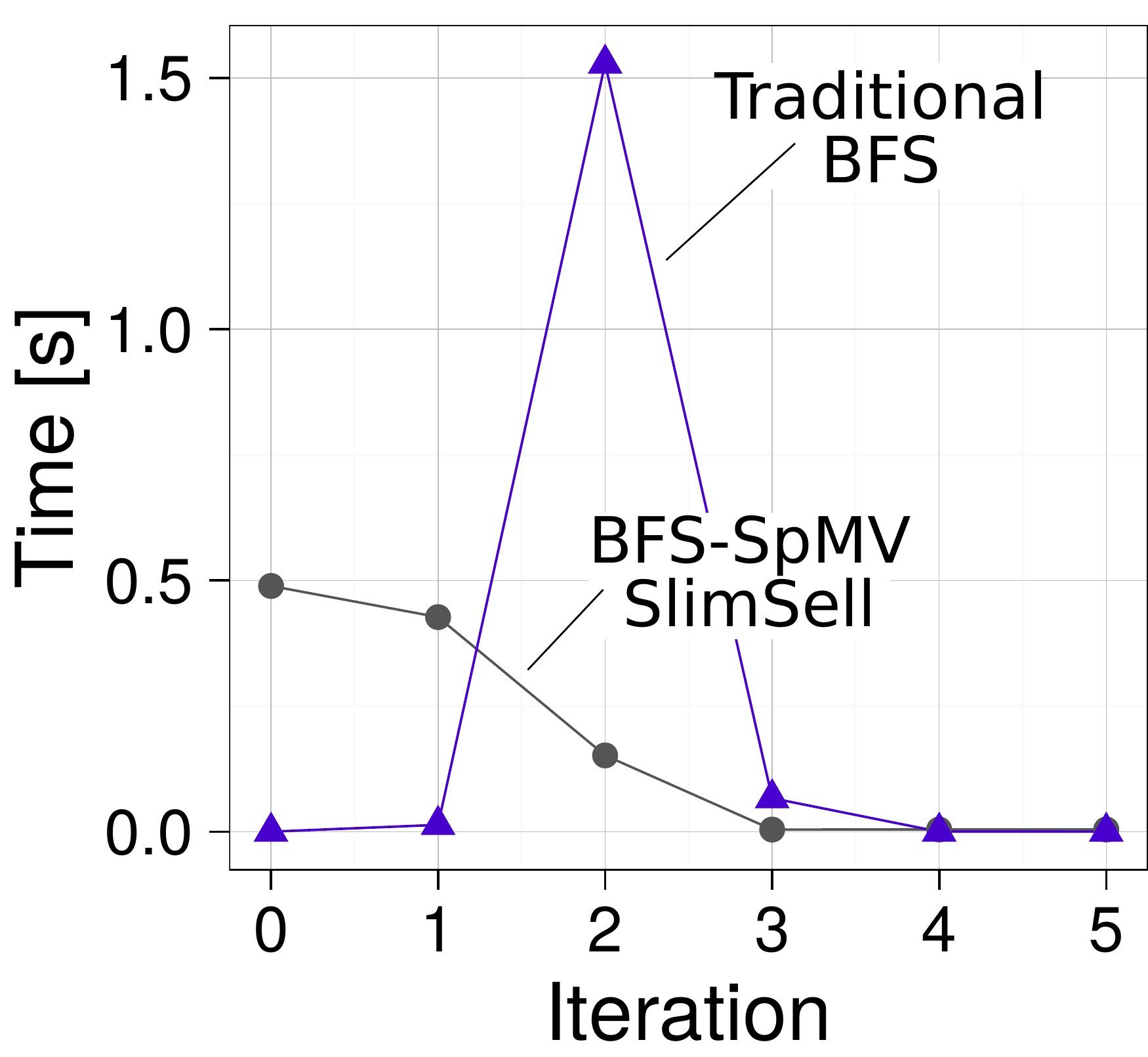}
        \caption{$n=2^{20}, \overline{\rho} = 512$}
    \end{subfigure}
    \begin{subfigure}[t]{0.2 \textwidth}
        \centering
        \includegraphics[width=\textwidth]{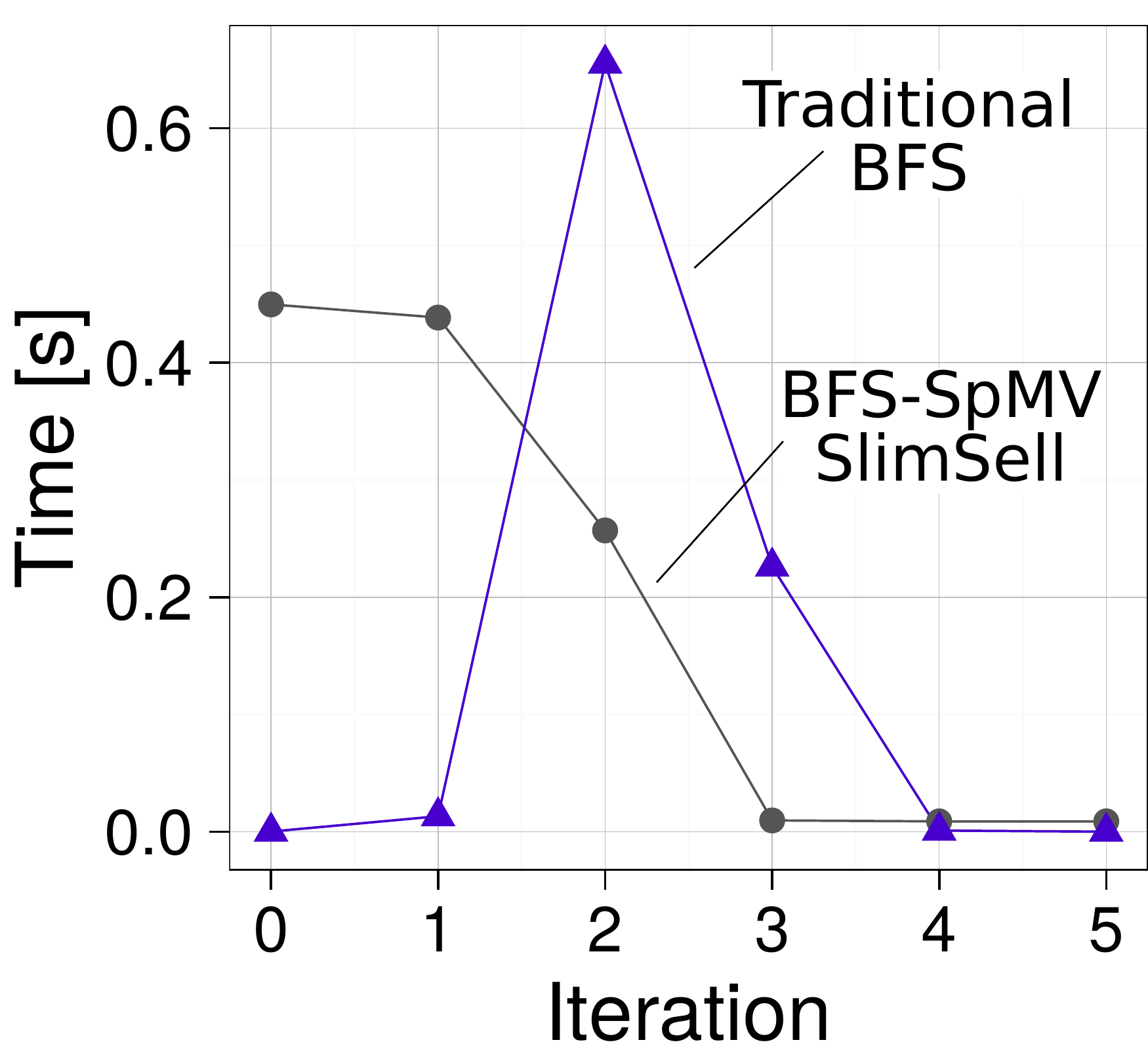}
        \caption{$n=2^{21}, \overline{\rho} = 128$}
    \end{subfigure}
    \caption{(\cref{sec:knl}, KNL analysis) Fine-grained comparison between \texttt{BFS-Trad} and \texttt{BFS-SpMV} with SlimSell and sel-max; $C$ is $16$, $G$ is Kronecker.} \label{fig:knl-fine}
\end{figure*}

\subsection{Manycore Analysis (KNL)}
\label{sec:knl}

Here, we evaluate SlimSell on Intel KNLs; see Figure~\ref{fig:knl}. The latency of each
iteration grows as expected with the increasing $\overline{\rho}$ or $n$.
KNL secures a drop in compute time after the first
iteration. In GPUs, 
this effect is less visible
due to larger SIMD widths that 
hinder reducing computation.
On CPUs, it is only negligibly visible in Figure~4d as
Erdős-Rényi graphs are used there.
%
%
Later (\cref{sec:state}) we show that our variant of BFS-SpMV based on SlimSell
outperforms the traditional work-efficient BFS on KNLs.


\subsection{Preprocessing Analysis}
\label{sec:size}

We illustrate that the full sorting time required for good storage
improvements as well as the actual build time can be amortized away. 
%
%
For a Kronecker graph with $n=2^{24}$,
sorting takes $\approx$0.95s, which constitutes $\approx$21\% of a single BFS run.
%
%
Thus, 10 BFS runs are enough to reduce the sorting time to $<$2\% of the total runtime.
The build time is analogous to that of Sell-$C$-$\sigma$~\cite{DBLP:journals/corr/KreutzerHWFB13}; it lasts longer but can also
be amortized with more BFS runs.
For example, on a Kronecker graph with $n=2^{18}$,
%
%
20 BFS runs are enough to reduce the full preprocessing time to $<$5\% of the total runtime.
The results for other graphs are similar.
In our analyses, we amortize the preprocessing time and report averaged iteration or total execution times.
%
%

\subsection{Storage Analysis}
\label{sec:size}

Figure~\ref{fig:size} illustrates storage improvements of SlimSell for various $\sigma, n, \overline{\rho}$, and graph families.
SlimSell reduces the size of
Sell-$C$-$\sigma$ by $\approx$50\%, a consistent advantage for various $n$,
$\bar{\rho}$, and $\sigma$. Notably, for full sorting ($\sigma = n$) and for Kronecker graphs, SlimSell is
also more compact than AL (by $\approx$5-10\%) which is effectively the
smallest graph representation if no compression is used. The same effect sets
in for $\sigma \ge \sqrt{n}$ for RW graphs.
Note that $C=8$ in the above analysis. Larger $C$ brings even more storage reductions.


\subsection{Comparison to the State-of-the-Art}
\label{sec:state}

\begin{figure*}[t]
    \centering
    %
    \begin{subfigure}[t]{0.2 \textwidth}
        \centering
        \includegraphics[width=\textwidth]{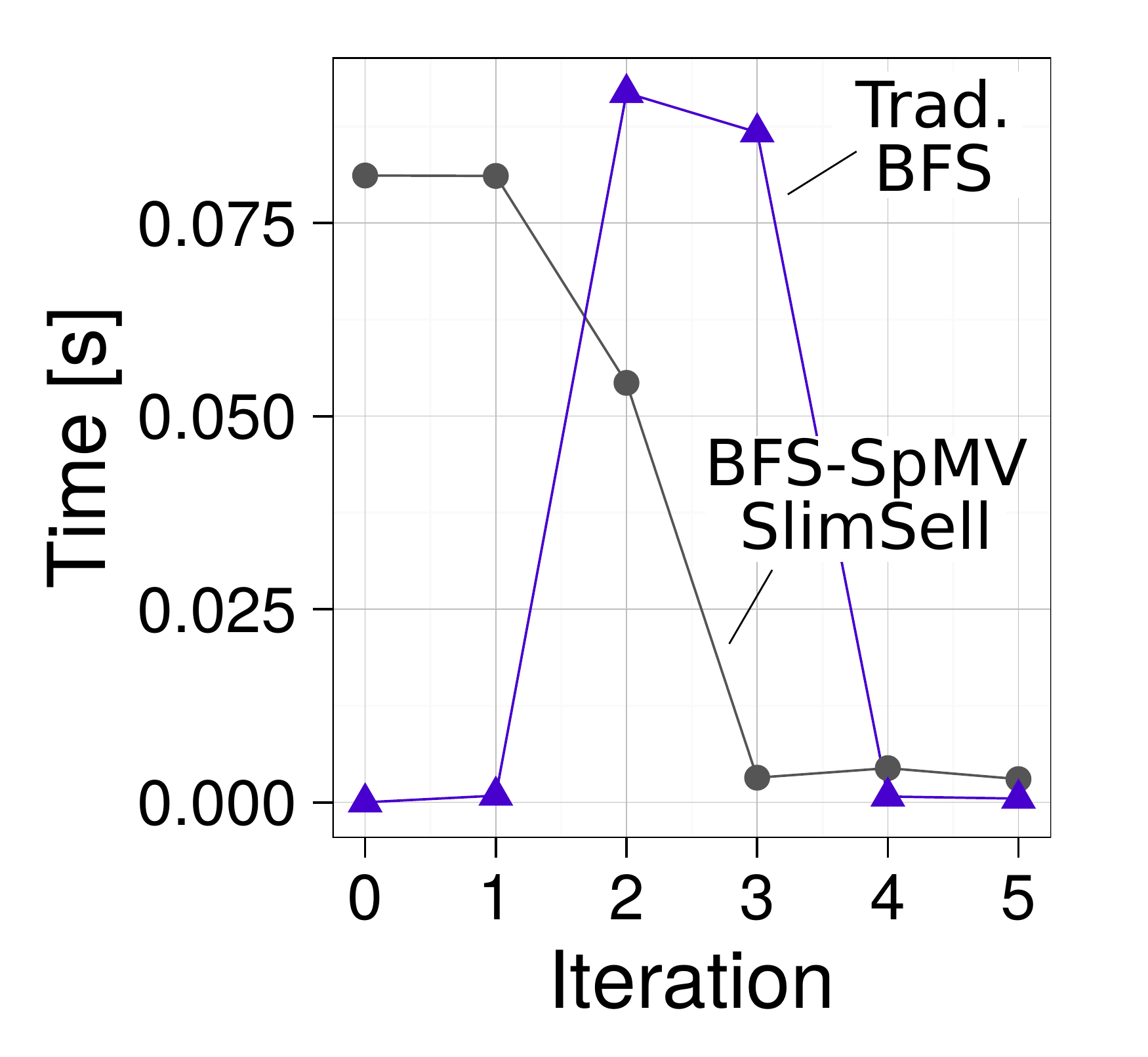}
        \caption{$\overline{\rho} = 128$}
    \end{subfigure}
    \hspace{-1em}
    \begin{subfigure}[t]{0.2 \textwidth}
        \centering
        \includegraphics[width=\textwidth]{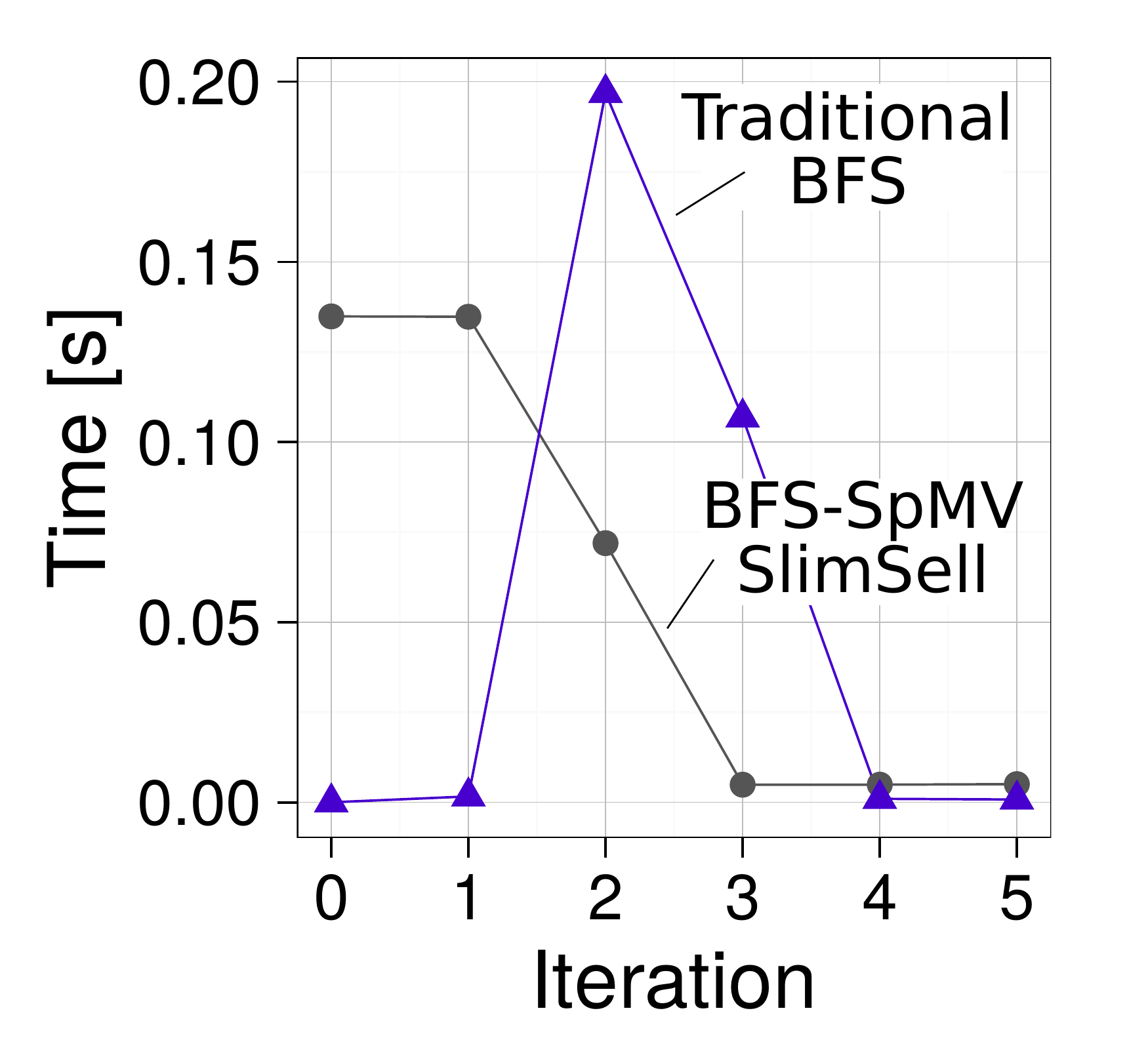}
        \caption{$\overline{\rho} = 256$}
    \end{subfigure}
    \hspace{-1em}
    \begin{subfigure}[t]{0.2 \textwidth}
        \centering
        \includegraphics[width=\textwidth]{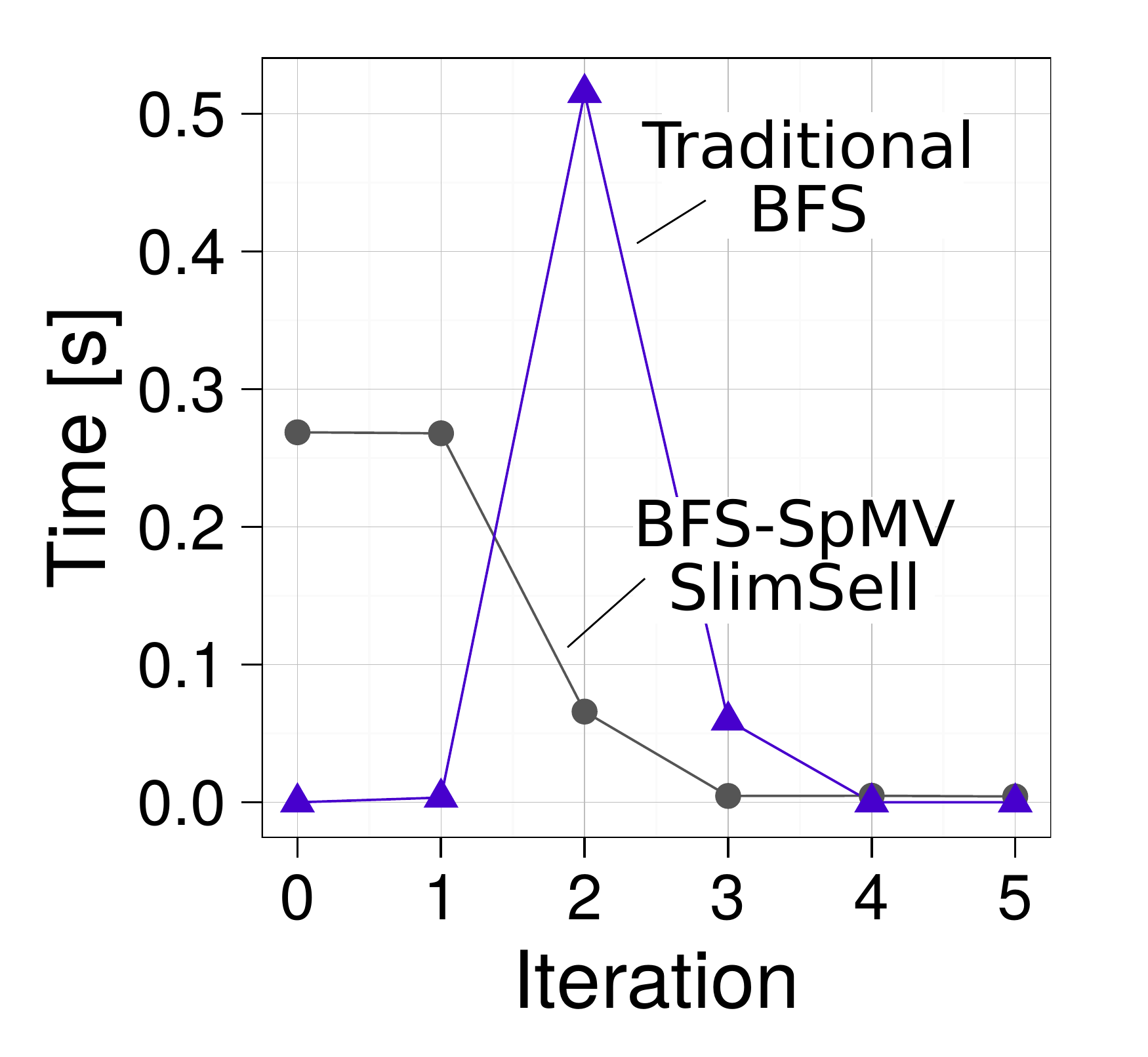}
        \caption{$\overline{\rho} = 512$}
    \end{subfigure}
    %
    \caption{(\cref{sec:state}) Fine-grained comparison between \texttt{BFS-Trad} (optimized for CPUs) and \texttt{BFS-SpMV} executing on GPUs based on the tropical semiring; $C$ is $32$, $G$ is Kronecker, $n = 2^{20}$.} \label{fig:state}
\end{figure*}

We compare BFS-SpMV equipped with SlimSell to \texttt{Trad-BFS} in a fine-grained analysis.
%
%
%
Figure~\ref{fig:state} shows the time to finish each BFS iteration for various
$\overline{\rho} \in \{1, ..., 512\}$; we compare the optimized traditional BFS running on
a system where it achieves highest performance (Xeon CPU) with BFS-SpMV based on SlimSell
that executes on Tesla K80 GPU. The higher $\overline{\rho}$ (denser $G$), the faster \texttt{BFS-SpMV} is. This is supported with the
intuition: more edges in $G$ entail a larger SIMD potential.
%
%
%
Finally, Figures~\ref{fig:motivate} and~\ref{fig:knl-fine} show that BFS-SpMV based
on SlimSell outperforms the traditional BFS by up to 53\%; denser graphs entail larger speedups.
As expected, power-law graphs are more SIMD-friendly during the expansion of
the large frontier part. Contrarily,  Erdős-Rényi graphs do not have this
property.
We conclude that SlimSell enables BFS-SpMV 
(running on throughput-oriented architectures such as GPUs)
to match the performance of highly-optimized work-efficient codes (executing on optimized
and latency-oriented CPUs).

\section{Related Work}


Extensive research has been done to address the challenges in 
parallel graph processing~\cite{lumsdaine2007challenges}.
One category of works includes parallel graph processing engines (e.g.,
Pregel~\cite{malewicz2010pregel} or PBGL~\cite{gregor2005parallel}) and various
improvements for algorithms and implementations 
%
%
such as direction-inversion~\cite{harish2007accelerating, suzumura2011performance}).
\sethlcolor{white}\hl{The well-known
direction-optimization}~\cite{beamer2013direction} \hl{and other work-avoidance schemes are orthogonal to
our work and can be implemented on top of SlimSell; see Figure~1.}\sethlcolor{white}
In general, we do not focus on improving the well-studied queue-based
BFS algorithm, but instead apply \emph{changes to the
graph representation} to improve performance.


There are several works on vectorization and GPUs for
BFS~\cite{paredes2016breadth, merrill2015high, cheng2014understanding,
yang2015fast, deng2009taming, harish2007accelerating}. Yet, they use
queue-based approaches or sparse matrix-sparse vector products and
thus suffer from the common downsides such as bad data locality, irregular data
accesses, and SIMD or vectorization
underutilisation~\cite{cheng2014understanding}. Contrarily, we start from
sparse-matrix dense-vector products as a basis for SlimSell that come with
regular memory accesses and better SIMD utilization.
%
%
\sethlcolor{white}\hl{The SlimSell optimization for unweighted graphs is
applicable not only to Sell-$C$-$\sigma$ but also other sparse matrix formats
such as ELLPACK/ELL, sliced ELLPACK~}\cite{monakov2010automatically}\hl{, ESB~}\cite{liu2013efficient}\sethlcolor{white}\hl{, ACSR~}\cite{Ashari:2014:FSM:2683593.2683679}\hl{, and CSR. They
all address general matrices and thus use the $val$ array with matrix values that can be omitted
to reduce data transfer.
Next, 
SlimSell removes padding and thus storage overheads inherent in ELLPACK,
differs from ELLPACK/ELL and ESB as it turns the graph storage layout 
by 90 degrees in memory along with Sell-$C$-$\sigma$ to utilize more of SIMD
compute power, and targets not only GPUs (like sliced ELLPACK or ACSR) or only KNL (like ESB),
but can be used for both and for CPUs.
Finally, our work is the first to conduct a systematic analysis on
using different semirings for BFS and on work complexity of BFS for miscellaneous
graph classes and schemes.
} 
%


%

  SlimChunk is similar to ACSR~\cite{Ashari:2014:FSM:2683593.2683679} and to CSR-VectorL in CSR-Adaptive~\cite{daga2015structural}. Unlike them, we do not adaptively split chunks
  and instead let the dynamic nature of the partial chunk allocation deal with work imbalance.
The advantage of our approach lies in its simplicity: no complex adaptive
schemes are required. In addition,
SlimChunk is by default enabled only for GPUs as this is the only architecture
that entailed load imbalance. Thus, SlimSell offers an intuitive and
simple design for CPUs and Xeon Phi.
Another similar scheme is SELL-P~\cite{Anzt:2015:AGK:2879157.2879167}: it 
pads the rows with zeros to make the row length of each block 
a multiple of the number of threads assigned to a row. Yet, this is discussed only
for GPU-based Krylov solvers.
%
%
Furthermore, Yang et al.~\cite{yang2011fast} facilitate matrix tiling to enhance temporal locality,
  a scheme different from SlimSell that uses Sell-$C$-$\sigma$ to turn the matrix layout by 90 degrees for better
  SIMD utilization, reduces work and storage overheads to limit the pressure on the memory subsystem.
Next, Gunrock~\cite{wang2016gunrock} is a CUDA library for graph-processing on
GPUs. It does not prescribe a high-performance graph representation, but instead provides a high-level, bulk-synchronous, data-centric abstraction 
that allows programmers to quickly develop new graph primitives. SlimSell could 
substitute graph representations used by default
in Gunrock for higher speedups; an idea that we leave for future work.
Another thread of future work is to port SlimSell to massively threaded systems such as SPARC64 or Adapteva Parallel.

Some work has been done into the formulation and implementation of BFS based on
the algebraic formulation.
%
%
Examples include GraphBLAS~\cite{mattson2014standards}, Combinatorial BLAS
(CombBLAS)~\cite{bulucc2011combinatorial}, and BFS with 2D adjacency data
distribution~\cite{bulucc2011parallel}. 
%
%
SlimSell, SlimWork, and
SlimChunk can be used to accelerate CombBLAS. Finally, our first systematic 
discussion and theoretical analysis of BFS based on various semirings may be
incorporated into the theoretical GraphBLAS framework.



\section{Conclusion}


Vectorization and throughput-oriented GPUs are reshaping graph
processing: traditional BFS codes optimized for latency-oriented CPUs will be
inefficient on systems with wide SIMD units.
We address this trend and propose SlimSell: a vectorizable graph representation
for BFS based on sparse matrix-dense vector products.  
  SlimSell augments the state-of-the-art Sell-$C$-$\sigma$
  matrix storage format along three dimensions: size, work, and load balance.
  It effectively
  applies the regular SIMD parallelism to irregular graph traversals to match
  or outperform in various scenarios 
  optimized Graph500 codes tuned for high-performance
  CPUs.  
  
  Furthermore, our work provides the first systematic analysis of the
formulation, performance, and work/storage complexity of algebraic BFS variants
based on different semirings. We derive bounds for uniform and power-law
graphs; our theoretical insights combined with performance studies 
for state-of-the-art architectures such as Intel Knights Landing
can be used
by architects and developers to improve heuristics and graph representations
implemented in processing engines such as
CombBLAS~\cite{bulucc2011combinatorial}.

We strongly believe that SlimSell can be used to accelerate other graph
algorithms, for example schemes for solving Betweenness
Centrality~\cite{solomonik2017scaling}. Potential speedups can be much higher
because BFS is a data-driven scheme with different memory access patterns
across iterations while many algorithms (e.g., Pagerank) have identical
communication patterns in each superstep. Finally, the principles behind
SlimSell could be extended to distributed-memory
settings~\cite{besta2015accelerating, besta2015active, besta2014fault,
gerstenberger2013enabling, besta2014slim}.  We leave these directions for
future work.

\maciej{From Edgar: One new thought I have is that maybe
it would be interesting to represent $\mathbf{f}_{i-1}$ block-sparse and with
the same block size as used for the Sell layout.}

\maciej{From Edgar: this does not seem to account for disambiguation of parents when they
are not unique. it also seems that you later do this on the fly while it could
be done all at once at the end (its completely parallel unlike BFS).}

\vspace{1em}



\macb{Acknowledgements: }
We thank Hussein Harake, Colin McMurtrie, and the whole CSCS team granting access to the Greina, Piz Dora, and Daint machines, and
for their excellent technical support. 
We thank Moritz Kreutzer and the anonymous reviewers for their insightful comments.





\bibliographystyle{abbrv}
\bibliography{references}

\maciej{From Edgar: One new thought I have is that maybe
it would be interesting to represent $\mathbf{f}_{i-1}$ block-sparse and with
the same block size as used for the Sell layout.}

\maciej{From Edgar: this does not seem to account for disambiguation of parents when they
are not unique. it also seems that you later do this on the fly while it could
be done all at once at the end (its completely parallel unlike BFS).}

\end{document}